%% file: paper.tex
\documentclass{article}
\usepackage{fullpage}

\title
 {A Pseudopolynomial Algorithm \\ for Alexandrov's Theorem%
   \thanks{A preliminary version of this paper appears in
     \emph{Proceedings of the 11th Algorithms and Data Structures
       Symposium}, Banff, Canada, 2009.}}

\usepackage{authblk}

\author[1]{%
  Daniel Kane%
  \thanks{Partially supported by an NDSEG Fellowship.}}
\author[2]{%
  Gregory N. Price%
  \thanks{Partially supported by an NSF Graduate Research Fellowship.}}
\author[2]{%
  Erik D. Demaine%
  \thanks{Partially supported by NSF CAREER award CCF-0347776.}}

\affil[1]{
  Department of Mathematics, Harvard University\\
  1 Oxford Street, Cambridge,~MA~02139, USA\\
  \protect\url{dankane@math.harvard.edu}}
\affil[2]{
  MIT Computer Science and Artificial Intelligence Laboratory\\
  32 Vassar Street, Cambridge,~MA~02139, USA\\
  \protect\url{{price,edemaine}@mit.edu}}

\date{}

\usepackage{amsthm}
\usepackage{amsmath}
\usepackage{amssymb}
\usepackage{url}
\usepackage{hyperref}

\begin{document}

\maketitle

\begin{abstract}
  \input{abstract}
\end{abstract}

\newcommand\wholelc{paper}
\newcommand\bigsec\section
\newcommand\bigref{Section}
\newcommand\biglc{section}
\newcommand\smallsec\subsection
\newcommand\smallref{Section}
\newcommand\smalllc{subsection}

\input{common}

\end{document}

%% file: abstract.tex
Alexandrov's Theorem states that every metric with the global topology
and local geometry required of a convex polyhedron is in fact the
intrinsic metric of a unique convex polyhedron.  Recent work by Bobenko
and Izmestiev describes a differential equation whose solution leads to the
polyhedron corresponding to a given metric.  We describe an algorithm
based on this differential equation to compute the polyhedron to
arbitrary precision given the metric, and prove a pseudopolynomial
bound on its running time.  Along the way, we develop pseudopolynomial
algorithms for computing shortest paths and weighted Delaunay triangulations
on a polyhedral surface, even when the surface edges are not shortest paths.

%% file: common.tex
\newtheorem{theorem}{Theorem}[\biglc]
\newtheorem{lemma}[theorem]{Lemma}
\newtheorem{proposition}[theorem]{Proposition}
\theoremstyle{definition}
\newtheorem{definition}[theorem]{Definition}

\newenvironment{idea}{\noindent{\it Proof idea.}}{}
\newcommand\term[1]{\emph{#1}}

\def\eps{\varepsilon}
\def\defeq{\stackrel\Delta=}

\newcommand{\matrixx}[4]%
{\scriptsize\begin{array}{@{}cc@{}}#1&#3\\#2&#4\end{array}}

\newcommand\R{\mathbb R}
\def\poly{\mathop{\rm poly}\nolimits}
\newcommand\per{\mathop{\rm per}}
\newcommand\area{\mathop{\rm area}}
\newcommand\vol{\mathop{\rm vol}}
\newcommand\argmin{\mathop{\rm arg\,min}}
\newcommand\argmax{\mathop{\rm arg\,max}}

\newcommand\tw[1]{\widetilde{#1}}

\newcommand\cprime{\/$'$}
\newcommand\J{\mathbf J}
\renewcommand\H{\mathbf H}
\newcommand\Fi{E_i}

\newcommand\algo\textsc
\newenvironment{algorithm}[1]
{\par\medskip\noindent\textbf{Algorithm} \algo{#1}.\medspace}
{\hfill$\square$}

\newcommand\comment[1]{}

\bigsec{Introduction}
\label{sec:introduction}

Alexandrov's celebrated theorem \cite{Alexandrov-1937,Alexandrov}
characterizes the metrics of convex polyhedra.
More precisely, a convex polyhedron in Euclidean 3-space,
viewed as a two-dimensional surface,
induces an \emph{intrinsic metric} on the (surface) points of the polyhedron:
the distance between two such points is the length of the shortest path
connecting them, restricted to lie along the polyhedron.
We can divorce the intrinsic metric from the extrinsic embedding in 3-space,
and Alexandrov's Theorem will tell us whether such an abstract metric
could have come from a convex polyhedron.

The intrinsic metric of a convex polyhedron has three obvious properties.
First, the metric is \emph{polyhedral}:
every point but finitely many exceptions (the vertices)
looks flat in the metric,
meaning that it has a neighborhood isometric to a flat disk.
Second, the metric is \emph{(locally) convex}: every circle of radius $r$
has circumference at most $2 \pi r$.
Finally, treated as a topological space,
the metric is homeomorphic to a 2-sphere.

Alexandrov's Theorem \cite{Alexandrov-1937,Alexandrov}
says that these three necessary conditions
are also sufficient: every convex polyhedral metric $M$ homeomorphic to
a sphere can be isometrically embedded as a convex polyhedron,
meaning that its induced intrinsic metric is exactly~$M$.
Furthermore, the convex polyhedron is unique up to isometry of 3-space
(an extension of Cauchy's Rigidity Theorem \cite{Cauchy,Steinitz-Rademacher-1976}).
Thus the extrinsic geometry can be reconstructed purely from the
intrinsic geometry.

Unfortunately, Alexandrov's proof is not constructive, suggesting an
algorithmic problem: given a convex polyhedral metric homeomorphic to
a sphere, find an isometric embedding as a convex polyhedron.
More precisely, the polyhedral metric can be specified by a complex
of triangles with specified edge lengths and adjacency between triangles.
It is easy to check that the given metric satisfies the three Alexandrov
conditions (one even follows from our input representation).
The goal is to find (approximate) coordinates for the vertices that
(approximately) satisfy the edge-length constraints and convexity.

One motivation for this problem is the problem of folding a given
polygon of paper into precisely the surface of a convex polyhedron.
There are efficient algorithms to find one or all gluings of a given polygon's
boundary to itself so that the resulting metric satisfies Alexandrov's
conditions \cite{DDLO,LO}.
These algorithms produce the desired convex polyhedral metrics homeomorphic
to spheres, knowing from Alexandrov's Theorem that they correspond to
actual 3D polyhedra that can be folded from the polygon of paper.
But without an algorithm for Alexandrov's Theorem, we do not know how to
compute these polyhedra.

Sabitov \cite{Sabitov02,Sabitov96a,Sabitov96b,Sabitov98,GFALOP}
showed how to enumerate all the isometric mappings of a polyhedral metric
as a polyhedron (not necessarily convex), which immediately leads to an
algorithm for Alexandrov's Theorem.
This algorithm has the distinction of being exact on a real RAM
supporting polynomial root finding, which can also be implemented on a binary
computer with a logarithmic dependence on the desired accuracy~$\eps$.
Unfortunately, the necessary polynomials have degree $2^{\Theta(m)}$
for a polyhedron with $m$ edges \cite{FP},
leading to an exponential running time.  Without the convexity constraint,
this exponentiality is unsurprising, because there can be exponentially
many isometric mappings and hence exponential output size.
But it is not known how to accelerate this algorithm in the convex case.

The desire for a practical algorithm for Alexandrov's Theorem,
``either a polynomial-time algorithm or a numerical approximation procedure'',
is posed as \cite[Open Problem 23.22]{GFALOP}.
While the polynomial-time challenge remains open (and perhaps unlikely),
we come close in this paper by attaining a pseudopolynomial-time algorithm.
Our work is based on recent progress on the other goal,
a numerical approximation procedure.

Namely, recent work by Bobenko and Izmestiev \cite{BI}
(building on work of Volkov and Podgornova \cite{Volkov-Podgornova-1971})
provides a new proof of Alexandrov's Theorem.
Their proof describes a certain ordinary differential equation (ODE)
and initial conditions whose solution contains sufficient information
to construct the embedding by elementary geometry.
The work in \cite{BI} was accompanied by a computer implementation of the
ODE~\cite{implementation}, which empirically produces accurate approximations
of embeddings of metrics on which it is tested.

We describe an algorithm based on the Bobenko--Izmestiev ODE,
and prove a pseudopolynomial bound on its running time.
Specifically, call an embedding of a convex polyhedral metric $M$
\term{$\eps$-accurate} if the metric is distorted by at most a factor
$1+\eps$, and \term{$\eps$-convex} if each dihedral
angle is at most $\pi + \eps$.
Then we show the following theorem:

\begin{theorem}\label{thm:main}
  Given a convex polyhedral metric $M$ homeomorphic to a sphere
  with $n$ vertices, ratio $S$ between the largest and smallest distance
  between vertices, and defect (discrete Gaussian curvature)
  between $\eps_1$ and $2\pi -
  \eps_8$ at each vertex, an $\eps_6$-accurate $\eps_9$-convex
  embedding of~$M$ can be found in time $
  \tw O\left(n^{915/2} S^{832}/(\eps^{121} \eps_1^{445} \eps_8^{617})
  \right)$ where $\eps = \min(\eps_6/nS, \eps_9\eps_1^2/nS^6)$.
\end{theorem}

The exponents in the time bound of Theorem~\ref{thm:main} are
remarkably large.  Thankfully, no evidence suggests that our algorithm
actually takes as long to run as the bound allows.  On the
contrary, our analysis relies on bounding approximately a dozen
geometric quantities, and to keep the
analysis tractable we use the simplest bound whenever available.
The algorithm's actual performance is governed by the
actual values of these quantities, and therefore by whatever sharper
bounds could be proved by a stingier analysis.

To describe our approach, consider an embedding of the metric $M$ as a
convex polyhedron in $\R^3$, and choose an arbitrary origin $O$ in the
surface's interior.  Then it is not hard to see that the $n$ distances
$r_i = \overline{Ov_i}$ from the origin to the vertices $v_i$,
together with $M$ and the combinatorial data describing which
polygons on $M$ are faces of the polyhedron, suffice to reconstruct
the embedding: the tetrahedron formed by $O$ and each triangle is
rigid in $\R^3$, and we have no choice in how to glue them to each
other.  In Lemma \ref{lem:9} below, we show that
in fact the radii alone suffice to reconstruct the embedding, to do so
efficiently, and to do so even with radii of finite precision.

Therefore in order to compute the unique embedding of $M$ that
Alexandrov's Theorem guarantees exists, we compute a set of radii $r =
\{r_i\}_i$ and derive a triangulation $T$.  The exact radii satisfy three
conditions:
\begin{enumerate}
\item the radii $r$ determine nondegenerate tetrahedra from $O$ to
  each face of $T$;
\item with these tetrahedra, the dihedral angles at each exterior edge
  total at most $\pi$; and
\item with these tetrahedra, the dihedral angles about each
  radius sum to $2\pi$.
\end{enumerate}
In our computation, we begin with a set of large initial radii $r_i =
R$ satisfying Conditions~1 and~2, and write $\kappa = \{\kappa_i\}_i$
for the differences by which Condition~3 fails about each radius.  We
then iteratively adjust the radii to bring $\kappa$ near zero and
satisfy Condition~3 approximately, maintaining Conditions~1 and~2
throughout.

The computation takes the following form.  We describe the Jacobian
$\left(\frac{\partial \kappa_i}{\partial r_j}\right)_{ij}$, showing
that it can be efficiently computed and that its inverse is
pseudopolynomially bounded.  We show further that the Hessian
$\left(\frac{\partial^2 \kappa_i}{\partial r_j\partial
    r_k}\right)_{ijk}$ is also pseudopolynomially bounded.  It follows
that a change in $r$ in the direction of smaller $\kappa$ as
described by the Jacobian, with some step size only pseudopolynomially
small, makes progress in reducing $|\kappa|$.  The step size can be
chosen online by doubling and halving, so it follows that we can take
steps of the appropriate size, pseudopolynomial in number, and obtain
an $r$ that zeroes $\kappa$ to the desired precision in
pseudopolynomial total time.  Theorem~\ref{thm:main} follows.

The construction of \cite{BI} is an ODE in the same $n$ variables
$r_i$, with a similar starting point and with the derivative of $r$
driven similarly by a desired path for $\kappa$.  Their proof
differs in that it need only show existence, not a bound, for the
Jacobian's inverse, in order to invoke the inverse function
theorem.  Similarly, while we must show a
pseudopolynomial lower bound (Lemma~\ref{lem:10}) on the altitudes of
the tetrahedra during our computation, the prior work shows
only that these altitudes remain positive.  In general our computation
requires that the known open conditions---this quantity is
positive, that map is nondegenerate---be replaced by stronger compact
conditions---this quantity is lower-bounded, that map's inverse is bounded.
We model our proofs of these strengthenings on the proofs in
\cite{BI} of the simpler open conditions, and we directly employ
several other results from that paper where possible.

One subroutine in our algorithm is of independent interest, which
computes the weighted Delaunay triangulation on a polyhedral surface.
Mitchell, Mount, and Papadimitriou \cite{MMP} solve the related problem of
computing the Voronoi diagram on a polyhedral surface.  However, their
algorithm assumes that the edges in the given triangulation of the surface are
shortest paths in the metric, which does not hold in our setting.
We show that their algorithm works for general triangulations, though
the running time increases from polynomial to pseudopolynomial.
Unfortunately, it seems difficult to dualize the weighted Voronoi
diagram and obtain a weighted Delaunay triangulation, because in the
weighted case, Voronoi cells can be empty and not contain their site.
Fortunately, we show that a dual transform is possible in the unweighted
case (even though Delaunay edges need not be shortest paths
between their endpoints), which lets us compute the unweighted Delaunay
triangulation.
Then we show that an incremental flip-based reweighting algorithm
lets us put the weights back into the problem,
while only requiring $O(n^2)$ flips and $O(n^2 \lg n)$ time.

The remainder of this \wholelc\ supplies the details of the proof of
Theorem~\ref{thm:main}.  We give background in
\bigref~\ref{sec:background}, and detail the main argument in
\bigref~\ref{sec:main}.  We bound the Jacobian in
\bigref~\ref{sec:jacobian} and the Hessian in
\bigref~\ref{sec:hessian}.  Some lemmas are deferred to
\bigref~\ref{sec:intermediate} for clarity.
Finally, \bigref~\ref{sec:triangulation} describes how to
compute weighted Delaunay triangulations on polyhedral surfaces.

\bigsec{Background and Notation}
\label{sec:background}

In this \biglc\ we define our major geometric objects
and give the basic facts about them.  We also define some parameters
describing our central object that we will need to keep bounded
throughout the computation.

\smallsec{Geometric notions}
Central to our argument are two dual classes of geometric structures
introduced by Bobenko and Izmestiev in~\cite{BI} under the names of
``generalized convex polytope'' and ``generalized convex polyhedron''.
Because in other usages the distinction between ``polyhedron'' and
``polytope'' is that a polyhedron is a three-dimensional polytope, and
because both of these objects are three-dimensional, we will refer to
these objects as ``generalized convex polyhedra'' and ``generalized
convex dual polyhedra'' respectively to avoid confusion.

First, we define the objects that our main theorem is about.

\begin{definition}
  A metric $M$ homeomorphic to the sphere is a \term{polyhedral
    metric} if each $x \in M$ has an open neighborhood isometric
  either to a subset of $\R^2$ or to a cone of angle less than $2\pi$
  with $x$ mapped to the apex.  The points falling into the latter
  case are called
  the \term{vertices} $V(M) = \{v_i\}_i$ of $M$, and they must be
  finite in number by compactness.

  The \term{defect} $\delta_i$ at a vertex $v_i \in V(M)$ is the
  difference between $2\pi$ and the total angle at the vertex, which
  is positive by the definition of a vertex.

  An \term{embedding} of $M$ is a piecewise linear map $f : M \to \R^3$.  An
  embedding $f$ is $\eps$-\term{accurate} if it distorts the metric
  $M$ by at most $1+\eps$, and $\eps$-\term{convex} if $f(M)$ is a
  polyhedron and each dihedral angle in $f(M)$ is at most $\pi +
  \eps$.

  A \term{perfect embedding} of a polyhedral metric $M$ is an isometry
  $f : M \to \R^3$ such that $f(M)$ is a convex polyhedron.
  Equivalently, an embedding is perfect if 0-accurate and 0-convex.
\end{definition}

Alexandrov's Theorem is that every polyhedral metric has a unique
perfect embedding, and our contribution is a pseudopolynomial-time
algorithm to construct $\eps$-accurate $\eps$-convex embeddings as
approximations to this perfect embedding.

\begin{definition}
  In a tetrahedron $ABCD$, write $\angle CABD$ for the dihedral angle
  along edge~$AB$.
\end{definition}

\begin{definition}
  A \term{triangulation} of a polyhedral metric $M$ is a decomposition
  into Euclidean triangles whose vertex set is $V(M)$.  Its vertices
  are denoted by $V(T) = V(M)$, its edges by $E(T)$, and its faces by
  $F(T)$.

  A \term{radius assignment} on a polyhedral metric $M$ is a map $r :
  V(M) \to \R_+$.  For brevity we write $r_i$ for $r(v_i)$.

  A \term{generalized convex polyhedron} is a gluing of metric
  tetrahedra with a common apex $O$.  The generalized convex
  polyhedron $P = (M, T, r)$ is determined by the polyhedral metric
  $M$ and triangulation $T$ giving its bases and the radius assignment
  $r$ for the side lengths.

  Write $\kappa_i \defeq 2\pi - \sum_{jk} \angle v_jOv_iv_k$ for the
  curvature about $Ov_i$, and $\phi_{ij} \defeq \angle v_iOv_j$ for
  the angle between vertices $v_i, v_j$ seen from the apex.
\end{definition}

Our algorithm, following the construction in~\cite{BI}, will choose a
radius assignment for the $M$ in question and iteratively adjust it
until the associated generalized convex polyhedron $P$ fits nearly
isometrically in $\R^3$.  The resulting radii will give an
$\eps$-accurate $\eps$-convex embedding of $M$ into $\R^3$.

In the argument we will require several geometric objects related to
generalized convex polyhedra.

\begin{definition}
  A \term{Euclidean simplicial complex} is a metric space on a
  simplicial complex where the metric restricted to each cell is
  Euclidean.

  A \term{generalized convex polygon} is a Euclidean simplicial
  2-complex homeomorphic to a disk, where all
  triangles have a common vertex $V$, the total angle at $V$ is no
  more than $2\pi$, and the total angle at each other vertex is no
  more than $\pi$.

  Given a generalized convex polyhedron $P = (M, T, r)$, the
  corresponding \term{generalized convex dual polyhedron} $D(P)$ is a
  certain Euclidean simplicial 3-complex.
  Let $O$ be a vertex called the \term{apex}, $A_i$ a vertex with
  $OA_i = h_i \defeq 1/r_i$ for each $i$.

  For each edge $v_iv_j \in E(T)$ bounding triangles $v_iv_jv_k$ and
  $v_jv_iv_l$, construct two simplices $OA_iA_{jil}A_{ijk}$,
  $OA_jA_{ijk}A_{jil}$ in $D(P)$ as follows.  Embed the two tetrahedra
  $Ov_iv_jv_k, Ov_jv_iv_l$ in $\R^3$.  For each $i' \in \{i, j, k,
  l\}$, place $A_{i'}$ along ray $Ov_{i'}$ at distance $h_{i'}$, and
  draw a perpendicular plane $P_{i'}$ through the ray at $A_{i'}$.
  Let $A_{ijk}, A_{jil}$ be the intersection of the planes $P_i, P_j,
  P_k$ and $P_j, P_i, P_l$ respectively.  By a standard computation in
  inversive geometry, $A_{ijk}$ and $A_{jil}$ are on the respective
  perpendicular rays from $O$ through $v_iv_jv_k$ and $v_jv_iv_l$, so
  $OA_iA_{jil}A_{ijk}$ and $OA_jA_{ijk}A_{jil}$ share the orientation
  of $Ov_iv_jv_k$ and $Ov_jv_iv_l$ because the two tetrahedra are
  together convex at edge $v_iv_j$ by local convexity.

  Now identify the vertices $A_{ijk}, A_{jki}, A_{kij}$ for each
  triangle $v_iv_jv_k \in F(T)$ to produce the Euclidean simplicial
  3-complex $D(P)$.  Since the six simplices produced
  about each of these vertices $A_{ijk}$ are all defined by the same
  three planes $P_i, P_j, P_k$ with the same relative configuration in
  $\R^3$, the total dihedral angle about each $OA_{ijk}$ is $2\pi$.
  On the other hand, the total dihedral angle about $OA_i$ is $2\pi -
  \kappa_i$, and the face about $A_i$ is a generalized convex polygon
  of defect $\kappa_i$.  Let
  $$ h_{ij} = \frac{h_j - h_i \cos\phi_{ij}}{\sin\phi_{ij}} $$
  be the altitude in this face from its apex $A_i$ to side $A_{ijk}A_{jil}$.
\end{definition}

\begin{definition}
  A \term{spherical simplicial 2-complex} is a metric space on a
  simplicial complex where each 2-cell is isometric to a spherical
  triangle.

  A \term{singular spherical polygon} (or \term{triangle},
  \term{quadrilateral}, etc) is a spherical simplicial 2-complex
  homeomorphic to a disk, where the total angle at each
  interior vertex is at most $2\pi$.  A singular spherical polygon is
  \term{convex} if the total angle at each boundary vertex is at most
  $\pi$.

  A \term{singular spherical metric} is a spherical simplicial
  2-complex homeomorphic to a sphere, where the total angle
  at each vertex is at most $2\pi$.
\end{definition}

The Jacobian bound in \bigref~\ref{sec:jacobian} makes use of certain
multilinear forms described in~\cite{BI}.

\begin{definition}\label{def:dual-stuff}
  The \term{dual volume} $\vol(h)$ is the volume of the generalized
  convex dual polyhedron $D(P)$, a cubic form in the dual altitudes
  $h$.

  The \term{mixed volume} $\vol(\cdot, \cdot, \cdot)$ is the symmetric
  trilinear form that formally extends the cubic form $\vol(\cdot)$:
  $$ \vol(a,b,c) \defeq
  \frac16 ( \vol(a+b+c) - \vol(a+b) - \vol(b+c) - \vol(c+a)
            + \vol(a) + \vol(b) + \vol(c) ).
  $$

  The $i$th \term{dual face area} $\Fi(g(i))$ is the area of the face
  around $A_i$ in $D(P)$, a quadratic form in the altitudes $g(i)
  \defeq \{h_{ij}\}_j$ within this face.

  The $i$th \term{mixed area} $\Fi(\cdot, \cdot)$ is the symmetric
  bilinear form that formally extends the quadratic form~$\Fi(\cdot)$:
  $$ \Fi(a, b) \defeq \frac12 ( \Fi(a+b) - \Fi(a) - \Fi(b) ) . $$

  Let $\pi_i$ be the linear map
  $$\pi_i(h)_j \defeq \frac{h_j - h_i \cos\phi_{ij}}{\sin\phi_{ij}} $$
  so that $\pi_i(h) = g(i)$.  Then define
  $$ F_i(a, b) \defeq \Fi(\pi_i(a), \pi_i(b)). $$
  so that $F_i(h, h) = \Fi(g(i), g(i))$ is the area of face $i$.
\end{definition}

By elementary geometry $\vol(h,h,h) = \frac13 \sum_i h_i F_i(h, h)$, so that by
a simple computation
$$ \vol(a, b, c) = \frac13 \sum_i a_i F_i(b, c). $$

\smallsec{Weighted Delaunay triangulations}
\label{subsec:weighted-delaunay}

The triangulations we require at each step of the computation are the
weighted Delaunay triangulations used in the construction of
\cite{BI}.  We give a simpler definition inspired by Definition 14 of
\cite{Glickenstein}.

\begin{definition}\label{def:delaunay}
  In a polyhedral metric $M$ with a radius assignment $r$, the
  \term{weight} of a vertex $v$ is the square of its radius, so that
  $h(v) = r(v)^2$.

  The \term{power} $\pi_v(p)$ of a point $p$ against a vertex $v$ in a
  polyhedral metric $M$ with weights $w$ is $pv^2 - w(v)$.

  The \term{center} $C(v_iv_jv_k)$ of a triangle $v_iv_jv_k \in T(M)$
  when embedded in $\R^2$ is the unique point $p$ such that
  $\pi_{v_i}(p) = \pi_{v_j}(p) = \pi_{v_k}(p)$, which exists by the
  radical axis theorem from classical geometry.  The quantity
  $\pi_{v_i}(p) = \pi(v_iv_jv_k)$ is the \term{power} of the triangle.

  A triangulation $T$ of a polyhedral metric $M$ with weights $w$ is
  \term{locally convex} at edge $v_iv_j$ with
  neighboring triangles $v_iv_jv_k, v_jv_iv_l$ if
  $\pi_{v_l}(C(v_iv_jv_k)) \geq \pi(v_iv_jv_k)$ and
  $\pi_{v_k}(C(v_jv_iv_l)) \geq \pi(v_jv_iv_l)$ when $v_iv_jv_k,
  v_jv_iv_l$ are embedded together in $\R^2$.  The two inequalities
  are equivalent by a lemma in classical geometry.  The triangulation
  is \term{strictly locally convex} at $v_iv_j$ if the inequalities
  hold strictly.

  A \term{weighted Delaunay triangulation} for vertex weights $w$ or
  radius assignment $r$
  on a polyhedral metric $M$ is a triangulation $T$ that is locally
  convex at every edge.
\end{definition}

We describe in \bigref~\ref{sec:triangulation} an algorithm
\algo{Polyhedral-Weighted-Delaunay} to compute a weighted Delaunay
triangulation in time $\tw O(n^3 S / \eps_8)$.

The radius assignment $r$ and triangulation $T$ admit a tetrahedron
$Ov_iv_jv_k$ just if the power of $v_iv_jv_k$ is negative, and the
squared altitude of $O$ in this tetrahedron is $-\pi(v_iv_jv_k)$.  The
edge $v_iv_j$ is convex when the two neighboring tetrahedra are
embedded in $\R^3$ just if it is locally convex in the triangulation
as in Definition~\ref{def:delaunay}.  A weighted Delaunay
triangulation with negative powers therefore gives a valid generalized
convex polyhedron if the curvatures $\kappa_i$ are positive.  For each
new radius assignment $r$ in the computation of \bigref~\ref{sec:main}
we therefore compute a weighted Delaunay triangulation and proceed
with the resulting generalized convex polyhedron, in which
Lemma~\ref{lem:10} guarantees a positive altitude and the choices in
the computation guarantee positive curvatures.

\smallsec{Notation for bounds}

\begin{definition}    
  Let the following bounds be observed:
  \begin{enumerate}
  \item $n$ is the number of vertices on $M$.  By Euler's formula,
    $|E(T)|$ and $|F(T)|$ are both $O(n)$.
  \item $\eps_1 \defeq \min_i \delta_i$ is the minimum defect.
  \item $\eps_2 \defeq \min_i (\delta_i - \kappa_i)$ is the minimum
    defect-curvature gap.
  \item $\eps_3 \defeq \min_{ij \in E(T)} \phi_{ij}$ is the minimum angle
    between radii.
  \item $\eps_4 \defeq \max_i \kappa_i$ is the maximum curvature.
  \item $\eps_5 \defeq \min_{v_iv_jv_k \in F(T)} \angle v_iv_jv_k$ is
    the smallest angle in the triangulation.  Observe that obtuse
    angles are also bounded: $\angle v_iv_jv_k < \pi - \angle
    v_jv_iv_k \leq \pi - \eps_5$.
  \item $\eps_6$ is used for the desired accuracy in embedding $M$.
  \item $\eps_7 \defeq (\max_i \frac{\kappa_i}{\delta_i})/(\min_i
    \frac{\kappa_i}{\delta_i}) - 1$ is the extent to which the ratio among
    the $\kappa_i$ varies from that among the $\delta_i$.  We will
    keep $\eps_7 < \eps_8 / 4\pi$ throughout the computation.
  \item $\eps_8 \defeq \min_i (2\pi - \delta_i)$ is the minimum angle around
    a vertex, the complement of the maximum defect.
  \item $\eps_9$ is used for the desired approximation to convexity in
    embedding $M$.
  \item $D$ is the diameter of $M$.
  \item $L$ is the maximum length of any edge in the input triangulation.
  \item $\ell$ is the shortest distance $v_iv_j$ between vertices.
  \item $S \defeq \max(D,L)/\ell$ is the maximum ratio of distances.
  \item $d_0 \defeq \min_{p \in M} Op$ is the minimum height of the apex off
    of any point on $M$.
  \item $d_1 \defeq \min_{v_iv_j \in E(T)} d(O,v_iv_j)$ is the minimum
    distance from the apex to any edge of $T$.
  \item $d_2 \defeq \min_i r_i$ is the minimum distance from the apex to
    any vertex of $M$.
  \item $H \defeq 1/d_0$; the name is justified by $h_i = 1/r_i \leq
    1/d_0$.
  \item $R \defeq \max_i r_i$, so $1/H \leq r_i \leq R$ for all $i$.
  \item $T \defeq HR$ is the maximum ratio of radii.
  \end{enumerate}
\end{definition}

Of these bounds, $n, \eps_1, \eps_8,$ and $S$ are fundamental to the given
metric $M$ or the form in which it is presented as input,
and $D, L,$ and $\ell$ are dimensionful parameters of the same metric input.  The values
$\eps_6$ and $\eps_9$ define the objective to be achieved, and our
computation will drive $\eps_4$ toward zero while maintaining $\eps_2$
large and $\eps_7$ small.  In \bigref~\ref{sec:intermediate} we bound
the remaining parameters $\eps_3, \eps_5, R, d_0, d_1,$ and $d_2$ in terms of these.

\begin{definition}
  Let $\J$ denote the Jacobian $\big(\frac{\partial \kappa_i}{\partial
    r_j}\big)_{ij}$, and $\H$ the Hessian $\big(\frac{\partial
    \kappa_i}{\partial r_j \partial r_k}\big)_{ijk}$.
\end{definition}

\bigsec{Main Theorem}
\label{sec:main}

In this \biglc, we prove our main theorem using the results proved in
the remaining \biglc s.  Recall

\medskip

{\bf Theorem \ref{thm:main}.}  {\it
  Given a polyhedral metric $M$ with $n$ vertices, ratio $S$ (the
  \term{spread}) between the diameter and the smallest distance
  between vertices, and defect at least $\eps_1$ and at most $2\pi -
  \eps_8$ at each vertex, an $\eps_6$-accurate $\eps_9$-convex
  embedding of $M$ can be found in time $
  \tw O\left(n^{915/2} S^{832}/(\eps^{121} \eps_1^{445} \eps_8^{617})
  \right)$ where $\eps = \min(\eps_6/nS, \eps_9\eps_1^2/nS^6)$.
}

\medskip

The algorithm of Theorem~\ref{thm:main} obtains an approximate
embedding of the polyhedral metric $M$ in $\R^3$.  Its main
subroutine is described by the following theorem:

\begin{theorem}\label{thm:submain}
  Given a polyhedral metric $M$ with $n$ vertices, ratio $S$ (the
  \term{spread}) between the diameter and the smallest distance
  between vertices, and defect at least $\eps_1$ and at most $2\pi -
  \eps_8$ at each vertex, a radius assignment $r$ for $M$ with maximum
  curvature at most $\eps$ can be found in time $
  \tw O\left(n^{915/2} S^{832}/(\eps^{121} \eps_1^{445} \eps_8^{617})
  \right)$.
\end{theorem}

\begin{proof}
  Let a \term{good} assignment be a radius assignment $r$ that
  satisfies two bounds: $\eps_7 < \eps_8/4\pi$ so that
  Lemmas~\ref{lem:16}--\ref{lem:10} apply and $r$ therefore by the
  discussion in \smallref~\ref{subsec:weighted-delaunay} produces a
  valid generalized convex polyhedron for $M$, and $\eps_2 = \Omega(
  \eps_1^2\eps_8^3/n^2S^2)$ on which our other bounds rely.  By
  Lemma~\ref{lem:12}, there exists a good assignment $r^0$.  We will
  iteratively adjust $r^0$ through a sequence $r^t$ of good assignments
  to arrive at an assignment $r^N$ with maximum curvature
  $\eps_4^N < \eps$ as required.  At each step we recompute $T$ as a
  weighted Delaunay triangulation by algorithm
  \algo{Polyhedral-Weighted-Delaunay} of \bigref~\ref{sec:triangulation}.

  Given a good assignment $r = r^n$, we will compute another good assignment
  $r' = r^{n+1}$ with $\eps_4 - \eps_4'
  = \Omega\left(\eps_1^{445} \eps_4^{121} \eps_8^{616} /
    (n^{907/2} S^{831}) \right)$.
  It follows that from $r^0$ we can arrive at a satisfactory $r^N$ with
  $N
  = O\left((n^{907/2} S^{831})/(\eps^{121} \eps_1^{445} \eps_8^{616})
  \right).$

  To do this, let $\J$ be the Jacobian $(\frac{\partial
    \kappa_i}{\partial r_j})_{ij}$ and $\H$ the Hessian
  $\big(\frac{\partial \kappa_i}{\partial r_j \partial
    r_k}\big)_{ijk}$, evaluated at $r$.  The goodness conditions and
  the objective are all in terms of $\kappa$, so we choose a desired
  new curvature vector $\kappa^*$ in $\kappa$-space and apply the
  inverse Jacobian to get a new radius assignment $r' = r + \J^{-1}
  (\kappa^* - \kappa)$ in $r$-space.  The actual new curvature vector
  $\kappa'$ differs from $\kappa^*$ by an error at most $\frac12 |\H|
  |r'-r|^2 \leq \left(\frac12 |\H| |\J^{-1}|^2\right) |\kappa^* - \kappa|^2$,
  quadratic in the desired change in curvatures with a coefficient
  $$ C \defeq \frac12 |\H| |\J^{-1}|^2
  = O\left(\frac{n^{3/2}S^{14}}{\eps_5^3}\frac{R^{23}}{D^{14}d_0^3d_1^8}
           \left(\frac{n^{7/2}T^2}{\eps_2\eps_3^3\eps_4}R\right)^2
         \right)
  = O\left(\frac{n^{905/2} S^{831}}
                {\eps_1^{443} \eps_4^{121} \eps_8^{616}}
         \right)
  $$
  by Theorems~\ref{thm:jacobian} and~\ref{thm:hessian} and
  Lemmas~\ref{lem:5}, \ref{lem:12}, \ref{lem:10}, and~\ref{lem:2}.

  Therefore pick a step size $p$, and choose $\kappa^*$ such that
  \begin{equation} \label{eq:stepsize-kappa}
    \kappa^*_i - \kappa_i
    = - p \kappa_i
      - p \left(\kappa_i - \delta_i \min_j\frac{\kappa_j}{\delta_j}\right).
  \end{equation}
  Consider a hypothetical $r^*$ that gives the curvatures $\kappa^*$,
  and examine the conditions on $\eps_4, \eps_2, \eps_7$ in turn.

  Both terms on the right-hand side of (\ref{eq:stepsize-kappa}) are
  nonpositive, so each $\kappa_i$ decreases by at least
  $p\kappa_i$.  Therefore the maximum curvature $\eps_4$ 
  decreases by at least $p\eps_4$.  If any defect-curvature gap
  $\delta_i - \kappa_i$ is less than $\eps_1/2$, then it increases by
  at least $p\kappa_i \geq p(\delta_i - \eps_1/2) \geq p(\eps_1/2)$; so the
  minimum defect-curvature gap $\eps_2$ either increases by at least $p\eps_1/2$
  or is at least $\eps_1/2$ already.  Finally, the $-p\kappa_i$ term decreases
  each $\kappa_i$ in the same ratio and therefore preserves $\eps_7$,
  and the $- p
  \left(\kappa_i - \delta_i \min_j(\kappa_j/\delta_j)\right)$ term
  decreases each ratio $\kappa_i/\delta_i$ by $p$ times the difference
  $\left(\kappa_i/\delta_i - \min_j(\kappa_j/\delta_j)\right)$
  and therefore reduces $\eps_7$ by $p\eps_7$.  Therefore $\kappa^*$
  would satisfy all three conditions with some room to spare.

  In particular, if we choose $p$ to guarantee that each $\kappa_i'$
  differs from $\kappa^*_i$ by at most $p\eps_4/2$, at most
  $p\eps_1/2$, and at most $p(\eps_1/4\pi)(\min_i \kappa_i)$, then
  this discussion shows that the step from $r$ to $r'$ will make at
  least half the ideal progress $p\eps_4$ in
  $\eps_4$ and keep $\eps_2, \eps_7$ within bounds.

  Since $$\min_i \kappa_i
  \geq (\max_j \kappa_j) (\min_{ij}\delta_i/\delta_j)(1+\eps_7)^{-1}
  \geq \eps_4 (\eps_1/2\pi)/2 = \eps_1\eps_4/4\pi$$
  and since
  $$ |\kappa' - \kappa^*|_\infty
  \leq |\kappa' - \kappa^*|
  \leq C|\kappa^* - \kappa|^2
  \leq 4C p^2 |\kappa|^2
  \leq 4C p^2 n \eps_4^2
  $$
  this can be done by choosing
  \begin{equation}
    \label{eq:stepsize-p}
    p = \eps_1^2/64\pi^2 n \eps_4 C ,
  \end{equation}
  which produces a good radius assignment $r'$ in which $\eps_4$ has
  declined by at least
  $$
    \frac{p\eps_4}2
    = \frac{\eps_1^2}{128\pi^2 n C}
    = \Omega\left(\frac{\eps_1^{445} \eps_4^{121} \eps_8^{616}}
                       {n^{907/2} S^{831}}
           \right)
  $$
  as required.  Any smaller $p$ will also produce a good assignment
  $r'$ and decrease $\eps_4$ by at least $p \eps_4/2$ proportionally.

  As a simplification, we need not compute $p$ exactly according to
  (\ref{eq:stepsize-p}).  Rather, we choose the step size $p^t$ at
  each step, trying first $p^{t-1}$ (with $p^0$ an arbitrary constant)
  and computing the actual curvature error $|\kappa' - \kappa^*|$.
  If the error exceeds its maximum acceptable value $p\eps_1^2
  \eps_4/16\pi^2$ then we halve $p^t$ and try step $t$ again, and if
  it falls below half this value then we double $p^t$ for the next
  round.  Since we double at most once per step and halve at most once
  per doubling plus a logarithmic number of times to reach an
  acceptable $p$, this doubling and halving costs only a constant factor.
  Even more important than the resulting simplification of the algorithm,
  this technique holds out the hope of actual performance exceeding
  the proven bounds.

  Now each of the $N$ iterations of the computation go as follows.
  Compute a weighted Delaunay triangulation $T^t$ for $r^t$ in time
  $\tw O(n^3 S / \eps_8)$ by algorithm
  \algo{Polyhedral-Weighted-Delaunay}.  Compute the Jacobian
  $\J^t$ in time $O(n^2)$ using formulas (14, 15) in \cite{BI}.
  Choose a step size $p^t$, possibly adjusting it, as discussed above.
  Finally, take the resulting $r'$ as $r^{t+1}$ and continue.
  The computation of $\kappa^*$ to check $p^t$ runs in linear time,
  and that of $r'$ in time $O(n^\omega)$ where $\omega < 3$ is the
  time exponent of matrix multiplication.  Each iteration therefore
  costs time $\tw O(n^3 S / \eps_8)$, and the whole computation costs
  time $\tw O(N n^3 S / \eps_8)$
  as claimed.
\end{proof}

Now with our radius assignment $r$ for $M$ and the resulting
generalized convex polyhedron $P$ with curvatures all near zero, it
remains to approximately embed $P$ and therefore $M$ in $\R^3$.
To begin, we observe that this is easy to do given exact
values for $r$ and in a model with exact computation: after
triangulating, $P$ is made up of rigid tetrahedra and we embed one
tetrahedron arbitrarily, then embed each neighboring tetrahedron in turn.

In a realistic model, we compute only with bounded precision, and in
any case Theorem~\ref{thm:submain} gives us only curvatures near zero,
not equal to zero.  Lemma~\ref{lem:9} produces an embedding in this
case, settling for less than exact isometry and exact convexity.

\begin{lemma}\label{lem:9}
  There is an algorithm that, given a radius assignment $r$ for which
  the corresponding curvatures $\kappa_i$ are all less than $\eps =
  O\left(\min(\eps_6/nS, \eps_9\eps_1^2/nS^6)\right)$ for some
  constant factor, produces explicitly by vertex coordinates in time
  $\tw O(n^3 S / \eps_8)$ an $\eps_6$-accurate $\eps_9$-convex embedding of
  $M$.
\end{lemma}
\begin{proof}
  As in the exact case, triangulate $M$, embed one tetrahedron
  arbitrarily, then
  embed its neighbors successively.  Call the resulting configuration $Q$.
  The positive curvature will force gaps in $Q$ between the tetrahedra,
  but since the curvature around each radius is less
  than $\eps$, the several copies of each vertex will be separated by
  at most $n \eps D$.  Now replace the several copies of each vertex
  by their centroid, so that the tetrahedra are distorted but leave no
  gaps.  Call the resulting polyhedron $P$ and its surface metric
  $M'$.  The computation of a weighted Delaunay triangulation takes
  time $\tw O(n^3 S / \eps_8)$ by
  Algorithm~\algo{Polyhedral-Weighted-Delaunay}, and the remaining steps
  require time $O(n)$.  We claim this embedding is $\eps_6$-accurate
  and $\eps_9$-convex.

  To show $\eps_6$-accuracy, observe that since each copy of each
  vertex was moved by at most $n \eps D$ from $Q$ to $P$, no edge of
  any triangle was stretched by more than a ratio $n \eps S$, and the
  piecewise linear map between faces relates $M'$ to $M$ with
  distortion $n \eps S \leq \eps_6$ as required.

  Now we show $\eps_9$-convexity.  Consider two neighboring triangles
  $v_iv_jv_k, v_jv_iv_l$ in $T$; we will show the exterior dihedral
  angle is at least $-\eps_9$.  First, consider repeating the
  embedding with $Ov_iv_jv_k$ the original tetrahedron, so that
  $Ov_iv_jv_k, Ov_jv_iv_l$ embed without gaps.  This moves each vertex
  by at most $n\eps D$, and makes the angle $v_lv_iv_jv_k$ convex and
  the tetrahedron $v_lv_iv_jv_k$ have positive signed volume.  The
  volume of this tetrahedron in the $P$ configuration is therefore 
  at least $- n\eps D^3$, since the derivative of the volume in any
  vertex is the area of the opposite face, which is at always less
  than $D^2$ since the sides remain $(1+o(1))D$ in length.

  Therefore suppose the exterior angle $\angle v_lv_iv_jv_k$ is negative.
  Then by Lemma~\ref{lem:6} and Lemma~\ref{lem:2},
  $$ \sin \angle v_lv_iv_jv_k
  = \frac32 \frac{[v_lv_iv_jv_k][v_iv_j]}{[v_iv_jv_l][v_jv_iv_k]}
  \geq - \frac{(n\eps D^3)D}{(\ell^2\eps_5/4)^2}
  \geq - \eps \frac{576nS^6}{\eps_2^2}
  $$
  and since $\eps_2 \geq \eps_1/2$ at the end of the computation,
  $ \angle v_lv_iv_jv_k
  \geq - \eps 2304 n S^6 / \eps_1^2
  \geq - \eps_9
  $
  as claimed.
\end{proof}

We now have all the pieces to prove our main theorem.

\begin{proof}[Proof of Theorem~\ref{thm:main}]
  Let $\eps \defeq
  O\left(\min(\eps_6/nS, \eps_9\eps_1^2/nS^6)\right)$,
  and apply the algorithm of Theorem~\ref{thm:submain} to obtain in time $
  \tw O\left( n^{915/2} S^{832}/(\eps^{121} \eps_1^{445} \eps_8^{617}) \right)$
  a radius assignment $r$ for $M$ with maximum curvature
  $\eps_4 \leq \eps$.

  Now apply the algorithm of Lemma~\ref{lem:9} to obtain in time
  $O(n^3)$ the desired embedding and complete the computation.
\end{proof}

\bigsec{Bounding the Jacobian}
\label{sec:jacobian}

\begin{theorem}\label{thm:jacobian}
  The Jacobian $\J = \big(\frac{\partial \kappa_i}{\partial
    r_j}\big)_{ij}$ has inverse pseudopolynomially bounded by
  $|\J^{-1}| = O\big(\frac{n^{7/2}T^2}{\eps_2\eps_3^3\eps_4}R\big)$.
\end{theorem}
\begin{proof}
  Our argument parallels that of Corollary 2 in \cite{BI}, which
  concludes that the same Jacobian is nondegenerate.  Theorem~4 of
  \cite{BI} shows that this Jacobian equals the Hessian of the volume
  of the dual~$D(P)$.  The meat of the corollary's proof is in Theorem
  5 of \cite{BI}, which begins by equating this Hessian to the
  bilinear form $6\vol(h, \cdot, \cdot)$ derived from the mixed
  volume we defined in Definition~\ref{def:dual-stuff}.  So we
  have to bound the inverse of this bilinear form.

  To do this it suffices to show that the form $\vol(h, x, \cdot)$ has
  norm at least
  $\Omega\big(\frac{\eps_2\eps_3^3\eps_4}{n^{7/2}T^2} \frac{|x|}R\big)$
  for all vectors $x$.  Equivalently, suppose some $x$ has
  $\left|\vol(h,x,z)\right| \leq |z|$ for all $z$; we show $|x| =
  O\big(\frac{n^{7/2}T^2}{\eps_2\eps_3^3\eps_4}R\big)$.

  To do this we follow the proof in Theorem~5 of \cite{BI} that the
  same form $\vol(h, x, \cdot)$ is nonzero for $x$ nonzero.
  Throughout the argument we work in terms of the dual $D(P)$.

  Recall that for each $i$, $\pi_ix$ is defined as the
  vector $\{x_{ij}\}_j$.  It suffices to show that for all $i$
  $$ |\pi_i x|_2^2
  = O\left(\frac{n^3T^3}{\eps_2^2\eps_3\eps_4}R^2
           + \frac{n^2T^2}{\eps_2\eps_3\eps_4}R|x|_1\right) $$
  since then by Lemma~\ref{lem:15}
  $$ |x|_2^2 \leq \frac {4n} {\eps_3^2} \max_i |\pi_i x|_2^2
  = O\left(\frac{n^4T^3}{\eps_2^2\eps_3^3\eps_4}R^2
           + \frac{n^3T^2}{\eps_2\eps_3^3\eps_4}R|x|_1\right),
  $$
  and since $|x|_1 \leq \sqrt{n} |x|_2$ and $X^2 \leq a + bX$ implies
  $X \leq \sqrt a + b$,
  $ |x|_2 = O\left(\frac{n^{7/2}T^2}{\eps_2\eps_3^3\eps_4}R\right). $
  Therefore fix an arbitrary $i$, let $g = \pi_i h$ and $y = \pi_i x$,
  and we proceed to bound $|y|_2$.

  We break the space on which $\Fi$ acts into the 1-dimensional
  positive eigenspace of $\Fi$ and its $(k-1)$-dimensional negative
  eigenspace, since by Lemma 3.4 of \cite{BI} the signature of $\Fi$
  is $(1,k-1)$, where $k$ is the number of neighbors of $v_i$.  Write
  $\lambda_+$ for the positive eigenvalue and $-\Fi^-$ for the
  restriction to the negative eigenspace so that $\Fi^-$ is positive
  definite, and decompose $g = g_+ + g_-$, $y = y_+ + y_-$ by
  projection into these subspaces.  Then we have
  \begin{align*}
    G \defeq \Fi(g,g) &= \lambda_+ g_+^2 - \Fi^-(g_-,g_-)
                  \defeq \lambda_+ g_+^2 - G_- \\
             \Fi(g,y) &= \lambda_+ g_+ y_+ - \Fi^-(g_-,y_-) \\
    Y \defeq \Fi(y,y) &= \lambda_+ y_+^2 - \Fi^-(y_-,y_-)
                  \defeq \lambda_+ y_+^2 - Y_-
  \end{align*}
  and our task is to obtain an upper bound on $Y_- = \Fi^-(y_-,y_-)$,
  which will translate through our bound on the eigenvalues of $\Fi$
  away from zero into the desired bound on $|y|$.

  We begin by obtaining bounds on $|\Fi(g,y)|$, $G_-$, $G$, and $Y$.
  Since $|z| \geq \left|\vol(h,x,z)\right|$ for all $z$ and
  $\vol(h,x,z) = \sum_j z_j F_j(h,x)$, we have $$|\Fi(g, y)| =
  |F_i(h,x)| \leq 1.$$ Further,
  $\det\left(\matrixx{\Fi(g,g)}{\Fi(g,y)}{\Fi(y,g)}{\Fi(y,y)}\right) <
  0$ because $\Fi$ has signature $(1,1)$ restricted to the $(y,g)$
  plane, so by Lemma~\ref{lem:1} $$Y = \Fi(y, y)
  < \frac{R^2}{\eps_2}.$$ On the other hand $-|x|_1 < \sum_j x_j
  F_j(x,h) = \sum_j h_j F_j(x,x),$ so
  $$Y = \Fi(y, y) = F_i(x,x)
  > -\frac1{h_i}\left( (n-1)\frac{R^2}{\eps_2}H + |x|_1\right)
  > -\left( \frac{nT}{\eps_2} R^2 + R |x|_1\right).$$
  Now $G = \Fi(g,g) > 0$, being the area of the face about $A_i$ in
  $D(P)$.  We have $|\Fi| = O(n/\eps_3)$ by
  construction, so $G, G_- \leq G+G_- \leq |\Fi| |h|^2 =
  O(nH^2/\eps_3)$ and similarly $G = O(nH^2/\eps_3)$.  On the other
  hand we have $G = \Omega(\eps_2/R^2)$ by Lemma~\ref{lem:1}.

  Now, observe that $\lambda_+ y_+ g_+$ is the geometric mean
  $$\lambda_+ y_+ g_+ = \sqrt{(\lambda_+ g_+^2)(\lambda_+ y_+^2)}
  = \sqrt{(G + G_-)(Y + Y_-)}$$ and by Cauchy-Schwarz
  $\Fi^-(y_-,g_-) \leq \sqrt{G_- Y_-},$ so that
  \begin{multline*}
    1 \geq \Fi(y,g) \geq \sqrt{(G + G_-)(Y + Y_-)} - \sqrt{G_- Y_-} \\
    = \sqrt{Y_-} \frac{G}{\sqrt{G+G_-}+\sqrt{G_-}}
     +\sqrt{G+G_-} \frac{Y}{\sqrt{Y+Y_-}+\sqrt{Y_-}}.
  \end{multline*}
  If $Y \geq 0$, it follows that
  $$Y_- \leq \frac{2\sqrt{G+G_-}}{G}
  = O\left(\frac{n^2 T^2 R^2}{\eps_2^2\eps_3}\right).$$
  If $Y < 0$, then
  $$ 1 \geq \frac{G\sqrt{Y_-}}{2\sqrt{G+G_-}}
          - \frac{(-Y)\sqrt{G+G_-}}{\sqrt{Y_-}}
  $$
  so
  $$ Y_-
  \leq \frac{2\sqrt{G+G_-}}{G} \sqrt{Y_-}
       + \frac{2(-Y)(G+G_-)}{G},
  $$
  and because $X^2 \leq a + bX$ implies $X \leq \sqrt a + b$,
  $$ \sqrt{Y_-}
  \leq \frac{2\sqrt{G+G_-}}{G}
       + \frac{\sqrt{2(-Y)(G+G_-)}}{\sqrt G}
  $$
  so that
  $$ Y_- = O\left(\max\left(
      \frac{G+G_-}{G^2},
      (-Y)\frac{G+G_-}{G}
    \right)\right)
  = O\left(
      \frac{n^2T^3}{\eps_2^2\eps_3}R^2
      + \frac{nT^2}{\eps_2\eps_3} R |x|_1
  \right).
  $$

  In either case, using $Y \leq R/\eps_2^2$ and Lemma~\ref{lem:3}, we have
  $$ |y|_2^2 = y_+^2 + |y_-|_2^2
  \leq |\Fi^{-1}| \left((Y+Y_-) + Y_-\right)
  = O\left(\frac{n^3T^3}{\eps_2^2\eps_3\eps_4}R^2
           + \frac{n^2T^2}{\eps_2\eps_3\eps_4}R|x|_1\right)
  $$
  and the theorem follows.
\end{proof}

\begin{lemma}\label{lem:15}
  $|x|^2 \leq (4n/\eps_3^2) \max_i |\pi_i x|^2.$
\end{lemma}
\begin{proof}
  Let $i = \argmax_i |x_i|$, and let $v_j$ be a neighbor in $T$ of
  $v_i$.  Without loss of generality let $x_i > 0$.  Then
  $$ (\pi_j x)_i
  = \frac{x_i - x_j \cos \phi_{ij}}{\sin \phi_{ij}}
  \geq x_i \frac{1 - \cos \phi_{ij}}{\sin \phi_{ij}}
  = x_i \tan (\phi_{ij}/2)
  > x_i \phi_{ij} / 2
  \geq |x|_\infty \eps_3 / 2$$
  and it follows that
  $$ |\pi_i x|
  \geq |\pi_i x|_\infty
  > |x|_\infty \eps_3 / 2
  \geq |x| \eps_3 / 2 \sqrt n $$
  which proves the lemma.
\end{proof}

\begin{lemma}\label{lem:1}
  $F_i(h, h) > \eps_2 / R^2$.
\end{lemma}
\begin{proof}
  The proof of Proposition 8 in \cite{BI} shows that a certain
  singular spherical polygon has angular area $\delta_i - \kappa_i$,
  where the singular spherical polygon is obtained by stereographic
  projection of each simplex of $P_i^*$ onto a sphere of radius
  $1/r_i$ tangent to it.  The total area of the polygon is $(\delta_i
  - \kappa_i)/r_i^2$ at this radius, so because projection of a plane
  figure onto a tangent sphere only decreases area we have
  $F_i(h, h) = \area(P_i^*) > (\delta_i - \kappa_i)/r_i^2 > \eps_2/R^2$.
\end{proof}

\begin{lemma}\label{lem:3}
  The inverse of the form $\Fi$ is bounded by
  $|\Fi^{-1}| = O(n/\eps_4)$.
\end{lemma}
\begin{proof} We follow the argument in Lemma~3.4 of \cite{BI} that
  the same form is nondegenerate.  Let $\ell_j(y)$ be the length of
  the side between $A_i$ and $A_j$ in $D(P)$ when the altitudes
  $h_{ij}$ are given by $y$.  Since $\Fi(y) = \frac12 \sum_j \ell_j(y)
  y_j$ it follows that $\Fi(a, b) = \frac12 \sum_j \ell_j(a) b_j$.
  Therefore in order to bound the inverse of the form $\Fi$ it
  suffices to bound the inverse of the linear map~$\ell$.

  Consider a $y$ such that $|\ell(y)|_\infty \leq 1$; we will show
  $|y|_\infty = O(n/\eps_4)$.  Unfold the generalized polygon
  described by $y$ into the plane, apex at the origin; the sides are
  of length $\ell_j(y)$, so the first and last vertex are a distance
  at most $|\ell(y)|_1 \leq n$ from each other.  But the sum of the
  angles is at least $\eps_4$ short of $2\pi$, so this means all the
  vertices are within $O(n/\eps_4)$ of the origin;
  and the altitudes $y_j$ are no more than the distances from vertices
  to the origin, so they are also $O(n/\eps_4)$ as claimed.
\end{proof}

The proof of Theorem \ref{thm:jacobian} is complete.

\bigsec{Bounding the Hessian}
\label{sec:hessian}

In order to control the error in each step of our computation, we need
to keep the Jacobian $\J$ along the whole step close to the value it
started at, on which the step was based.  To do this we bound the
Hessian $\H$ when the triangulation is fixed, and we show that the
Jacobian does not change discontinuously when changing radii force a
new triangulation.

Each curvature $\kappa_i$ is of the form $2\pi - \sum_{j,k:v_iv_jv_k
  \in T} \angle v_jOv_iv_k$, so in analyzing its derivatives we focus
on the dihedral angles $\angle v_jOv_iv_k$.  When the tetrahedron
$Ov_iv_jv_k$ is embedded in $\R^3$, the angle $\angle v_jOv_iv_k$ is
determined by elementary geometry as a smooth function of the
distances among $O, v_i, v_j, v_k$.  For a given triangulation $T$
this makes $\kappa$ a smooth function of $r$.  Our first lemma shows
that no error is introduced at the transitions where the triangulation
$T(r)$ changes.

\begin{lemma}\label{lem:18}
  The Jacobian $\J = \big(\frac{\partial \kappa_i}{\partial
    r_j}\big)_{ij}$ is continuous at the boundary between radii
  corresponding to one triangulation and to another.
\end{lemma}
\begin{proof}
  Let $r$ be a radius assigment consistent with more than one
  triangulation, say with a flat face $v_iv_jv_kv_l$ that can be
  triangulated by $v_iv_k$ as $v_iv_jv_k, v_kv_lv_i$ or by $v_jv_l$ as
  $v_jv_kv_l, v_lv_iv_j$.  Since the Jacobian is continuous when
  either triangulation is fixed and $r$ varies, it suffices to show
  that for neighboring radius assigments $r + \Delta r$, the
  curvatures $\kappa$ obtained with either triangulation differ by a
  magnitude $O(|\Delta r|^2)$, with any coefficient determined by the
  polyhedral metric or the radius assigment $r$.

  Embed the two tetrahedra $Ov_iv_jv_k, Ov_kv_lv_i$ or $Ov_jv_kv_l,
  Ov_lv_iv_j$ together in $\R^3$, with distances $[Ov_i]$, etc., taken
  from $r + \Delta r$.  Of the ten pairwise distances between the five
  points in this diagram, eight are determined by $M$ or the radii and
  do not vary between the $v_iv_k$ and $v_jv_l$ diagrams.  Since the
  angles $\angle v_jOv_iv_k$, etc., are smooth functions of these ten
  distances, it suffices to show that the remaining two distances
  $[v_iv_k], [v_jv_l]$ differ between the diagrams by $O(|\Delta
  r|^2)$.  Letting $X$ denote the intersection of the geodesics
  $v_iv_k, v_jv_l$ on the face $v_iv_jv_kv_l$, we have $[v_iv_k]$ in
  the $v_iv_k$ diagram equal to $[v_iX]+[Xv_k]$, while in the $v_jv_l$
  diagram $v_iXv_k$ form a triangle with the same lengths $[v_iX],
  [Xv_k]$ and a shorter $[v_kv_k]$.  The difference between $[v_iv_k]$
  in the two diagrams is therefore the slack in the triangle
  inequality in this triangle $v_iXv_k$, which is bounded by
  $O(|\Delta r|^2)$ since the vertices have moved a distance
  $O(|\Delta r|)$ from where $r$ placed them with $v_i, X, v_k$
  collinear.
\end{proof}

It now remains to control the change in $\J$ as $r$ changes within any
particular triangulation, which we do by bounding the Hessian.

\begin{theorem}\label{thm:hessian}
  The Hessian $\H = \big(\frac{\partial
    \kappa_i}{\partial r_j \partial r_k}\big)_{ijk}$ is bounded in norm by
  $O\big(n^{5/2} S^{14} R^{23} / (\eps_5^3 d_0^3 d_1^8 D^{14})\big)$.
\end{theorem}
\begin{proof}
  It suffices to bound in absolute value each element
  $\frac{\partial^2 \kappa_i}{\partial r_j \partial r_k}$ of the
  Hessian.  Since $\kappa_i$ is $2\pi$ minus the sum of the dihedral
  angles about radius $r_i$, its derivatives decompose into sums of
  derivatives $\frac{\partial^2 \angle v_lOv_iv_m}{\partial
  r_j \partial r_k}$ where $v_iv_lv_m \in F(T)$.  Since the geometry
  of each tetrahedron $Ov_iv_lv_m$ is determined by its own side
  lengths, the only nonzero terms are where $j, k \in \{i, l, m\}$.

  It therefore suffices to bound the second partial derivatives of
  dihedral angle $AB$ in a tetrahedron $ABCD$ with respect to the
  lengths $AB, AC, AD$.  By Lemma~\ref{lem:8} below, these are
  degree-23 polynomials in the side lengths of $ABCD$, divided by
  $[ABCD]^3[ABC]^4[ABD]^4$.  Since $2[ABC], 2[ABD] \geq (D/S)d_1$,
  $6[ABCD] \geq d_0 (D/S)^2 \sin \eps_5$, and each side is $O(R)$, the
  second derivative is $O\big( S^{14} R^{23} / (\eps_5^3 d_0^3 d_1^8
  D^{14}) \big)$.

  Now each element in the Hessian is the sum of at most $n$ of these
  one-tetrahedron derivatives $\frac{\partial^2 \angle
  v_lOv_iv_m}{\partial r_j \partial r_k}$, and the norm of the Hessian
  itself is at most $n^{3/2}$ times the greatest absolute value of any
  of its elements, so the theorem is proved.
\end{proof}

\begin{definition}
  For the remainder of this \biglc, $ABCD$ is a tetrahedron and
  $\theta$ the dihedral angle $\angle CABD$ on $AB$.
\end{definition}

\begin{lemma}\label{lem:6}
  $$ \sin \theta = \frac32 \frac{[ABCD][AB]}{[ABC][ABD]}. $$
\end{lemma}
\begin{proof}
  First, translate $C$ and $D$ parallel to $AB$ to make $BCD$
  perpendicular to $AB$, which has no effect on either side of the
  equation.  Now $[ABCD] = [BCD][AB]/3$ while $[ABC] = [BC][AB]/2$ and
  $[ABD] = [BD][AB]/2$, so our equation's right-hand side is
  $\frac{2[BCD]}{[BC][BD]} = \sin \angle CBD = \sin \theta$.
\end{proof}

\begin{lemma}\label{lem:7}
  Each of the derivatives $\frac{\partial \theta}{\partial AB},
  \frac{\partial \theta}{\partial AC}, \frac{\partial \theta}{\partial
    AD}$ is a degree-10 polynomial in the side lengths of $ABCD$, divided by
  $[ABCD][ABC]^2[ABD]^2$.
\end{lemma}
\begin{proof}
  Write $[ABC]^2, [ABD]^2$ as polynomials in the side lengths using
  Heron's formula.  Write $[ABCD]^2$ as a polynomial in the side
  lengths as follows.  We have
  $ 36[ABCD]^2 = \det([\vec{AB}, \vec{AC}, \vec{AD}])^2 = \det(M) $
  where $M = [\vec{AB}, \vec{AC}, \vec{AD}]^T [\vec{AB}, \vec{AC},
  \vec{AD}]$.  The entries of $M$ are of the form $\vec u \cdot
  \vec v = \frac12 (|\vec u|^2 + |\vec v|^2 - |\vec u - \vec v|^2)$,
  which are polynomials in the side lengths.  With Lemma~\ref{lem:6},
  this gives $\sin^2 \theta$ as a rational function of the side
  lengths.

  Now $\frac{\partial \theta}{\partial x} = \frac{\partial \sin
      \theta}{\partial x} \big/ \sqrt{1-\sin^2 \theta}$ for any variable
  $x$, so the square of this first derivative is a rational function.
  Computing it in SAGE~\cite{sage} or another computer algebra system
  finds that for each $x \in \{AB, AC, AD\}$, this squared derivative
  has numerator the square of a degree-10 polynomial with denominator
  $[ABCD]^2[ABC]^4[ABD]^4$.  The lemma is proved.
\end{proof}

\begin{lemma}\label{lem:8}
  Each of the six second partial derivatives of $\theta$ in $AB, AC,
  AD$ is a degree-23 polynomial in the side lengths of $ABCD$, divided
  by $[ABCD]^3[ABC]^4[ABD]^4$.
\end{lemma}
\begin{proof}
  By Lemma~\ref{lem:7}, each first partial derivative is a degree-10
  polynomial divided by $[ABCD][ABC]^2[ABD]^2$.  Since $[ABCD]^2,
  [ABC]^2, [ABD]^2$ are polynomials of degree 6, 4, 4 respectively,
  their logarithmic derivatives have themselves in the denominator and
  polynomials of degree 5, 3, 3 respectively in the numerator.  The
  second partial derivatives therefore have an additional factor of
  $[ABCD]^2[ABC]^2[ABD]^2$ in the denominator and an additional degree of 13
  in the numerator, proving the lemma.
\end{proof}

\bigsec{Intermediate Bounds}
\label{sec:intermediate}

In this \biglc\ we bound miscellaneous parameters in the computation
in terms of the fundamental parameters $n, S,
\eps_1, \eps_8$ and the computation-driving parameter $\eps_4$.

\smallsec{Initial conditions}

\begin{lemma}\label{lem:12}
  Given a polyhedral metric space $M$, there exists a radius assignment $r$
  with curvature skew $\eps_7 < \eps_8 / 4\pi$, maximum radius $R =
  O(nD/\eps_1\eps_8)$, and minimum defect-curvature gap $\eps_2 =
  \Omega(\eps_1^2\eps_8^3/n^2S^2)$.
\end{lemma}

In the proof of Lemma~\ref{lem:12} we require a lemma from singular
spherical geometry.

\begin{lemma}\label{lem:14}
  Let $C$ be a convex singular spherical $n$-gon with one interior
  vertex $v$ of defect $\kappa$ and each boundary vertex $v_i$ a
  distance $\alpha \leq vv_i \leq \beta \leq \pi/2$ from $v$.  Then
  the perimeter $\per(C)$ is bounded by
  $$2\pi - \kappa - 2n(\pi/2-\alpha)
  \leq \per(C)
  \leq (2\pi - \kappa) \sin \beta. $$
\end{lemma}
\begin{proof}
  Embed $C$ in the singular spherical polygon $B$ that results from
  removing a wedge of angle $\kappa$ from a hemisphere.

  To derive the lower bound, let the nearest point on the equator to
  each $v_i$ be $u_i$, so that $u_iv_i \leq \pi/2-\alpha$.  Then by the
  triangle inequality,
  $$ \per(C) = \sum_{ij} v_iv_j \geq \sum_{ij} u_iu_j - v_iu_i -
  u_jv_j \geq 2\pi - \kappa - 2n(\pi/2-\alpha) .$$

  For the upper bound, let $D$ be the singular spherical surface
  obtained as the $\beta$-disk about $v$ in $B$.  Then $C$ can
  be obtained by cutting $D$ in turn along the geodesic extension of
  each of the sides of $C$.  Each of these cuts, because it is a
  geodesic, is the shortest path with its winding number and is
  therefore shorter than the boundary it replaces, so the perimeter
  only decreases in this process.  Therefore
  $ \per(C) \leq \per(D) = (2\pi - \kappa) \sin \beta .$
\end{proof}

\begin{proof}[Proof of Lemma~\ref{lem:12}]
  Let $r$ have the same value $R$ on all vertices.  We show that
  for sufficiently large $R = O(nD/\eps_1\eps_8)$ the assignment $r$
  is valid and satisfies the required bounds on $\eps_2$ and $\eps_7$.
  To do this it suffices to show that $\eps_2 \leq \delta_i - \kappa_i
  \leq \eps_7 \eps_1$ for the desired $\eps_2, \eps_7$ and each $i$.

  For each vertex $v_i$, consider the singular spherical polygon $C$
  formed at $v_i$ by the neighboring tetrahedra $v_iOv_jv_k$.  Polygon
  $C$ has one interior vertex at $v_iO$ with defect $\kappa_i$, its
  perimeter is $\sum_{jk} \angle v_jv_iv_k = 2\pi - \delta_i$, and
  each vertex $v_iv_k$ is convex.  The spherical distance from the
  center $v_iO$ to each vertex $v_iv_k$ is $\angle Ov_iv_k = \pi/2 -
  \Theta(v_iv_k/R)$, which is at least $\rho_{\min} \defeq \pi/2 -
  \Theta(D/R)$ and at most $\rho_{\max} \defeq \pi/2 - \Theta(\ell/R)$.
  Now by Lemma~\ref{lem:14} above, we have
  $$ 2\pi - \kappa_i  - 2n(\pi/2 - \rho_{\min})
  \leq 2\pi - \delta_i
  \leq (2\pi - \kappa_i) \sin \rho_{\max}. $$
  The left-hand inequality implies
  $$\delta_i - \kappa_i
  \leq 2n(\pi/2 - \rho_{\min})
  = O(n D/R)$$
  so that $\delta_i - \kappa_i \leq (\eps_8/4\pi)\eps_1$
  if $R = \Omega(nD/\eps_1\eps_8)$ for a sufficiently large
  constant factor.
  The right-hand inequality then implies
  $$ \delta_i - \kappa_i
  \geq (2\pi - \delta_i)\frac{1 - \sin \rho_{\max}}{\sin \rho_{\max}}
  \geq \eps_8 (1 - \sin \rho_{\max})
  = \Omega(\eps_8 \ell^2/R^2)
  = \Omega(\eps_1^2\eps_8^3 / n^2S^2)
  $$
  so that the $\eps_2$ bound holds.
\end{proof}

\smallsec{Two angle bounds}

\begin{lemma}\label{lem:5}
  $\eps_3 > \ell d_1 / R^2$.
\end{lemma}
\begin{proof}
  $\eps_3$ is the smallest angle $\phi_{ij}$ from the apex $O$ between
  any two vertices $v_iv_j$.  Now $v_iv_j \geq \ell$, and the altitude
  from $O$ to $v_iv_j$ is at least $d_1$.  Therefore $\frac12
  \ell d_1 \leq [Ov_iv_j] \leq \frac12 \sin \phi_{ij} R^2$, so $\phi_{ij} >
  \sin \phi_{ij} \geq \ell d_1 / R^2$.
\end{proof}

\begin{lemma}\label{lem:2}
  $\eps_5 > \eps_2/6S$.
\end{lemma}
\begin{proof}
  Suppose that a surface triangle has an angle of $\epsilon$; we want to show $\epsilon > \eps_2/6S$.
  Let the largest angle of that triangle be $\pi - \epsilon'$.  By the
  law of sines, $\frac{\sin\epsilon'}{\sin\epsilon} \leq S$, so
  $\epsilon > \sin\epsilon \geq \sin\epsilon' / S > \epsilon' / 3S$
  since $\epsilon' \leq 2\pi/3$ implies $\sin\epsilon'/\epsilon' > 1/3$.
  It therefore suffices to show that $\epsilon' \geq \eps_2 / 2$.

  Let the angle of size $\pi - \epsilon'$ be at vertex $i$.  Embed all
  of the tetrahedrons around $Ov_i$ in space so that all the faces
  line up except for the one corresponding to an edge $e$ adjacent to
  this angle of $\pi-\epsilon'$.  The two copies of $e$ are separated
  by an angle of $\kappa_i$.  Letting $f$ be the other side forming
  this large angle, the angle between one copy of $e$ and the copy of
  $f$ is $\pi-\epsilon'$.  Now the sum of all the angles around $v_i$
  is $2\pi - \delta_i$, so apply the triangle inequality for angles
  twice to deduce
\begin{align*}
\eps_2
& \leq 2\pi - (2\pi - \delta_i) - \kappa_i \\
& \leq 2\pi - ((\pi-\epsilon') + \angle f e') - \kappa_i \\
&    = \pi + \epsilon' - \angle f e' - \kappa_i \\
& \leq \pi + \epsilon' - ((\pi - \epsilon') - \kappa_i) - \kappa_i \\
&    = 2 \epsilon'.
\end{align*}
\end{proof}

\smallsec{Keeping away from the surface}
\label{subsec:wall-hitting}

In this section we bound $O$ away from the surface
$M$.  Recall that $d_2$ is the minimum distance from $O$ to any vertex
of $M$, $d_1$ is the minimum distance to any edge of $T$, and $d$ is
the minimum distance from $O$ to any point of $M$.

\begin{lemma}\label{lem:16}
  $d_2 = \Omega\big( D \eps_1\eps_4\eps_5^2\eps_8 / (nS^4) \big). $
\end{lemma}
\begin{proof}
  This is an effective version of Lemma 4.8 of \cite{BI}, on whose
  proof this one is based.

  Let $i = \argmin_i Ov_i$, so that $Ov_i = d_2$, and suppose that
  $d_2 = O\big(D \eps_1\eps_4\eps_5^2\eps_8 / (nS^4) \big)$ with a
  small constant factor.  We consider the singular spherical polygon
  $C$ formed at the apex $O$ by the tetrahedra about $Ov_i$.  First we
  show that $C$ is concave or nearly concave at each of its vertices,
  so that it satisfies the hypothesis of Lemma~\ref{lem:13}.  Then we
  apply Lemma~\ref{lem:13} and use the fact that the ratios of the
  $\kappa_j$ are within $\eps_7 \leq \eps_8/4\pi$ of those of the
  $\delta_j$ to get a contradiction.

  Consider a vertex of $C$, the ray $Ov_j$.  Let $v_iv_jv_k,
  v_jv_iv_l$ be the triangles in $T$ adjacent to $v_iv_j$, and embed
  the two tetrahedra $Ov_iv_jv_k, Ov_jv_iv_l$ in $\R^3$.  The angle
  of $C$ at $Ov_j$ is the dihedral angle $v_kOv_jv_l$.

  By convexity, the dihedral angle $v_kv_iv_jv_l$ contains $O$, so
  if $O$ is on the same side of plane $v_kv_jv_l$ as $v_i$ is then
  the dihedral angle $v_kOv_jv_l$ does not contain $v_i$ and is a
  reflex angle for $C$.  Otherwise, the distance from $O$ to this
  plane is at most $Ov_i = d_2$, and we will bound the magnitude of $\angle
  v_kOv_jv_l$.

  By Lemma~\ref{lem:6},
  $$ \sin \angle v_kOv_jv_l
  = \frac32 \frac{[Ov_kv_jv_l][Ov_j]}{[Ov_jv_k][Ov_jv_l]}.
  $$
  Now $[Ov_kv_jv_l] \leq d_2[v_kv_jv_l]/3 = O(d_2D^2)$ and $[Ov_j]
  \leq [Ov_i] + [v_iv_j] \leq D + d_2$.  On the other hand $[Ov_jv_k] =
  (1/2) [Ov_j][Ov_k] \sin \angle v_jOv_k$, and $[Ov_j], [Ov_k] \geq
  \ell - d_2$ while $\angle v_kv_iv_j \leq \angle v_iv_kO + \angle
  v_kOv_j + \angle Ov_jv_i \leq \angle v_kOv_j + O(d_2/D)$ so that
  $\angle v_jOv_k \geq \eps_5 - O(d_2/D)$, so $[Ov_jv_k] =
  \Omega(\ell^2 \eps_5)$, and similarly $[Ov_jv_l]$.  Therefore
  $ \sin \angle v_kOv_jv_l
  = O\left(d_2D^3 / (\ell^4 \eps_5^2)\right)
  = O\left(\eps_1\eps_4\eps_8/n\right), $
  and the angle of $C$ at $Ov_i$ is
  $$\angle v_kOv_jv_l = O\left(\eps_1\eps_4\eps_8/n\right). $$

  On the other hand observe that $\per(C) = \sum_{jk, v_iv_jv_k \in F(T)} \angle
  v_jOv_k \leq \sum_{jk} (\angle v_jv_iv_k + O(d_2/D)) = 2\pi -
  \delta_i + O(nd_2/D)$.

  Now apply Lemma~\ref{lem:13} to deduce that
  $$ \kappa_i + O(\eps_1\eps_4\eps_8)
  \geq \left(1 - \frac{\per(C)}{2\pi}\right) \sum_{j \neq i} \kappa_j
  \geq \left(\frac{\delta_i}{2\pi} - O(nd_2/D)\right) \sum_{j \neq i} \kappa_j
  $$
  so that
  \begin{align*}
  \frac{\kappa_i}{\delta_i} + O(\eps_4\eps_8)
  &\geq   \left(\frac 1 {2\pi} - O(nd_2/\eps_1D)\right)
          \sum_{j \neq i} \kappa_j \\
  &\geq   (1 + o(\eps_8)) \frac 1 {2\pi}
          \left(\min_j \frac{\kappa_j}{\delta_j}\right)
          \sum_{j \neq i} \delta_j \\
  &   =   (1 + o(\eps_8)) \frac{4\pi - \delta_i}{2\pi}
          \left(\min_j \frac{\kappa_j}{\delta_j}\right) \\
  &\geq   (1 + o(\eps_8)) (1 + \eps_8/2\pi)
          \left(\min_j \frac{\kappa_j}{\delta_j}\right)
  \end{align*}
  so that since $\kappa_i/\delta_i = \Omega(\eps_4)$,
  $$
  \frac{\kappa_i}{\delta_i}
  \left(\min_j \frac{\kappa_j}{\delta_j}\right)^{-1}
  \geq   (1 + O(\eps_8))^{-1} (1 + \eps_8/2\pi)
  $$
  which for a small enough constant factor on $d_2$ and hence on the
  $O(\eps_8)$ term makes $\eps_7 > \eps_8/(4\pi)$, which is a
  contradiction.
\end{proof}

\begin{lemma}\label{lem:17}
  $d_1 = \Omega(\eps_4^2\eps_5^2d_2^2/DS^2)
  = \Omega\big(D \eps_1^2\eps_4^4\eps_5^6\eps_8^2 / (n^2S^{10}) \big).$
\end{lemma}
\begin{proof}
  This is an effective version of Lemma 4.6 of \cite{BI}, on whose
  proof this one is based.

  Let $O$ be distance $d_1$ from edge $v_iv_j$, which neighbors faces
  $v_iv_jv_k, v_jv_iv_l \in F(T)$.  Consider the spherical
  quadrilateral $D$ formed at $O$ by the two tetrahedra $Ov_iv_jv_k,
  Ov_jv_iv_l,$ and the singular spherical quadrilateral $C$ formed by
  all the other tetrahedra.  We will show the perimeter of $C$ is
  nearly $2\pi$ for small $d_1$ and apply Lemma~\ref{lem:11} to deduce a
  bound.  This requires also upper and lower bounds on the side
  lengths of $C$ and a lower bound on its exterior angles.

  In triangle $Ov_iv_j$, let the altitude from $O$ have foot $q$; then
  $Oq = d_1$ while $v_iO, v_jO \geq d_2$, so $\angle v_jv_iq, \angle
  v_iv_jq = O(d_1/d_2)$.  Also, $qv_i, qv_j \geq d_2-d_1$, so $q$ is
  at least distance $(d_2-d_1)\sin \eps_5$ from any of $v_iv_k,
  v_kv_j, v_jv_l, v_lv_i$, and $O$ is at least $(d_2-d_1)\sin \eps_5 -
  d_1 = \Omega(d_2\eps_5)$ from each of these sides.

  Now $\angle v_iOv_j = \pi - O(d_1/d_2)$ is the distance on the
  sphere between opposite vertices $Ov_i, Ov_j$ of $D$, so by the
  triangle inequality the perimeter of $D$ is at least $2\pi -
  O(d_1/d_2)$.  Each side of~$C$ is at least $\Omega(\eps_5)$ and at
  most $\pi - \Omega(\eps_5d_2/D)$.

  In spherical quadrilateral $D$, the two opposite angles $\angle
  v_kOv_iv_l$, $\angle v_lOv_jv_k$ are each within $O(d_1/\eps_5d_2)$
  of the convex $\angle v_kv_iv_jv_l$ and therefore either reflex for
  $D$ or else at least $\pi - O(d_1/\eps_5d_2)$.  To bound the other
  two angles $\angle v_iOv_lv_j, \angle v_jOv_kv_i$, let the smaller
  of these be $\theta$; then by Lemma~\ref{lem:6},
  $$ \pi - \theta = O(\sin \theta)
  = O\left(\frac{(D^2d_1)D}{(\eps_5d_2D/S)^2}\right)
  = O\left(\frac{S^2Dd_1}{\eps_5^2d_2^2}\right). $$

  Now there are two cases.  In one case, $d_1 =
  \Omega(\eps_4\eps_5^2d_2^2/DS^2)$.  In the alternative, we find that
  each angle of $D$ is at least $\pi - \eps_4/2$ and each angle of~$C$
  at most $\pi - \eps_4/2$.  In the latter case applying
  Lemma~\ref{lem:11} to $C$ finds that $2\pi - O(d_1/d_2) = 2\pi -
  \Omega(\eps_4^2\eps_5d_2/D)$ so that $d_1 =
  \Omega(\eps_4^2\eps_5d_2^2/D)$.

  In either case $d_1 = \Omega(\min(\eps_4\eps_5^2d_2^2/DS^2,
  \eps_4^2\eps_5d_2^2/D)) = \Omega(\eps_4^2\eps_5^2d_2^2/DS^2)$, and
  the bound on $d_2$ from Lemma~\ref{lem:16} finishes the proof.
\end{proof}

\begin{lemma}\label{lem:10}
  $$d_0 =
  \Omega\left(\min\left(d_1 \sqrt{\eps_5} \eps_4,
  \frac{d_1^{3/2}\eps_4}{\sqrt D},
  \frac{d_1^2 \eps_4}{DS^2}\right)\right)
  = \Omega\left(\frac{\eps_1^4\eps_4^9\eps_5^{12}\eps_8^4}{n^4S^{22}} D\right)
  . $$
\end{lemma}
\begin{proof}
  This is an effective version of Lemma 4.5 of \cite{BI}, on whose
  proof this one is based.

  Let $O$ be distance $d_0$ from triangle $v_iv_jv_k \in F(T)$.
  Consider the singular spherical polygon $C$ cut out at $O$ by all
  the tetrahedra other than $Ov_iv_jv_k$.  We show lower and upper
  bounds on the side lengths of $C$ and lower bounds on its exterior
  angles, show the perimeter $\per(C)$ is near $2\pi$ for small $d_0$,
  and apply Lemma~\ref{lem:11} to derive a bound.

  The perimeter of $C$ is the total angle about $O$ on the faces of
  the tetrahedron $Ov_iv_jv_k$, which is $2\pi - O(d_0^2/d_1^2)$.  Each
  side of $C$ is at least $\Omega(\eps_5)$ and at most $\pi -
  \Omega(d_1/D)$.

  Let $\theta$ be the smallest dihedral angle of $\angle v_iOv_jv_k,
  \angle v_jOv_kv_i, \angle v_kOv_iv_j$.  Then by Lemma~\ref{lem:6},
  $$ \pi - \theta = O(\sin \theta)
  = O\left(\frac{(D^2d_0)D}{(d_1D/S)^2}\right)
  = O\left(\frac{S^2Dd_0}{d_1^2}\right). $$

  Now there are two cases.  If $\theta \leq \pi - \eps_4/2$, then it
  follows immediately that $d_0 = \Omega(d_1^2\eps_4/(S^2D))$.
  Otherwise, $\theta > \pi - \eps_4/2$, so the interior angles of $C$
  are more than $\eps_4/2$.  Applying Lemma~\ref{lem:11}, the perimeter
  $\per(C)$ is at most $2\pi - \Omega(\min(\eps_4^2\eps_5,
  \eps_4^2d_1/D))$, so that $d_0 = \Omega(\min(d_1\eps_4\eps_5^{1/2},
  d_1^{3/2}\eps_4D^{-1/2}))$.  The bound on $d_1$ from
  Lemma~\ref{lem:17} finishes the proof.
\end{proof}

\smallsec{Lemmas in spherical geometry}

These lemmas about singular spherical polygons and metrics are used in
\smallref~\ref{subsec:wall-hitting} above.

\begin{lemma}\label{lem:11}
  Let a convex singular spherical polygon have all exterior angles at
  least $\gamma$ and all side lengths between $c$ and $2\pi - c$.
  Then its perimeter is at most $2\pi - \Omega(\gamma^2 c)$.
\end{lemma}
\begin{proof}
  This is an effective version of Lemma~5.4 on pages~45--46
  of~\cite{BI}, and we follow their proof.  The proof in~\cite{BI}
  shows that the perimeter is in general bounded by the perimeter in
  the nonsingular case.  In this case consider any edge $AB$ of the
  polygon, and observe that since the polygon is contained in the
  triangle $ABC$ with exterior angles $\gamma$ at $A, B$ its perimeter
  is bounded by this triangle's perimeter.  Since $c \leq AB \leq 2\pi
  - c$, the bound follows by straightforward spherical geometry.
\end{proof}

\begin{lemma}\label{lem:13}
  Let $S$ be a singular spherical metric with vertices $\{v_i\}_i$,
  and let $C$ be the singular spherical polygon consisting of the
  triangles about some distinguished vertex $v_0$.  Suppose $C$ has
  $k$ convex vertices, each with an interior angle at least $\pi -
  \eps$ for some $\eps > 0$ and an exterior angle no more than $\pi$.
  Then
  $$ \kappa_0 + 2\eps k \geq \left(1 - \frac{\per(C)}{2\pi}\right)
  \sum_{i \neq 0} \kappa_i. $$
\end{lemma}
\begin{proof}
  We reduce to Lemma 5.5 from \cite{BI} by induction.  If $k=0$, so
  that all vertices of $C$ have interior angle at least $\pi$, then
  our statement is precisely theirs.

  Otherwise, let $v_i$ be a vertex of $C$ with interior angle $\pi -
  \theta \in [\pi - \eps, \pi)$.  Draw the geodesic from $v_i$ to
  $v_0$, and insert along this geodesic a pair of spherical triangles
  each with angle $\theta/2$ at $v_i$ and angle $\kappa_0/2$ at $v_0$,
  meeting at a common vertex $v_0'$.  The polygon $C'$ and
  triangulation $S'$ that result from adding these two triangles
  satisfy all the same conditions but with $k-1$ convex vertices on
  $C'$, so
  $$ \kappa_0' + 2\eps(k-1) \geq \left(1 - \frac{\per(C')}{2\pi}\right)
  \sum_{i \neq 0} \kappa_i'. $$
  Now $C'$ and $C$ have the same perimeter, $\kappa_0' \leq \kappa_0 +
  \theta \leq \kappa_0 + \eps$, $\kappa_i' = \kappa_i - \theta \geq
  \kappa_i - \eps$, and $\kappa_j' = \kappa_j$ for $j \not\in \{0,
  i\}$, so it follows that
  $$ \kappa_0 + 2\eps k
  \geq \kappa_0' + (2k-1)\eps
  \geq \left(1 - \frac{\per(C')}{2\pi}\right) \sum_{i \neq 0} \kappa_i $$
  as claimed.
\end{proof}

\bigsec[Delaunay triangulation]{Delaunay Triangulation on a Polyhedral Surface}
\label{sec:triangulation}

\def\axis{\mathop{\rm axis}\nolimits}

In this \biglc\ we present an algorithm
\algo{Polyhedral-Weighted-Delaunay} to compute a weighted Delaunay
triangulation on a polyhedral surface, as defined in
\smallref~\ref{subsec:weighted-delaunay} above, and we prove its
correctness and efficiency.
Our algorithm consists of three main parts.
First, in \smallref~\ref{subsec:shortest-paths},
we compute the unweighted Voronoi diagram on a polyhedral surface,
using a generalization of the Mitchell--Mount--Papadimitriou algorithm
to allow the surface to have non-shortest-path edges.
Second, in \smallref~\ref{subsec:unweighted-Delaunay},
we use this Voronoi diagram to compute an unweighted Delaunay triangulation,
which is complicated by the fact that there are many possible paths between
two vertices.
Finally, in \smallref~\ref{subsec:weighted-Delaunay-algo},
we show how to modify this triangulation into a weighted Delaunay
triangulation, by continuously reweighting the vertices and performing
flips as necessary.

\smallsec{Shortest paths on a non-shortest-path triangulation}
\label{subsec:shortest-paths}

In this \smalllc\ we describe modifications to the analysis of the ``continuous
Dijkstra'' algorithm of \cite{MMP} and \cite{Mou85}, which can be used
to compute an (unweighted) Voronoi diagram on a polyhedral surface.  Our
modifications permit the
algorithm to dispense with the assumption that the input triangulation
consists of shortest paths, at the cost of a modest loss in efficiency
when the assumption is not satisfied.

We first sketch the main ideas of \cite{Mou85} and \cite{MMP}.  These
papers describe an algorithm for computing shortest paths on the
surface of a polyhedron from a number of sources.  The shortest paths
on the surface are represented by the shortest paths to each edge,
approaching through each adjacent face.  Specifically, each edge, for
each adjacent face, is partitioned into intervals on which the
shortest paths to each point through that face originate at the same
source and pass through the same sequence of edges and vertices.  Then
for each such interval the algorithm considers the last vertex in the
sequence and records the location of that vertex in an edge unfolding
along the sequence of edges.  It is easy to see that once this
representation is obtained, standard planar techniques suffice to
efficiently obtain the shortest path to any point and the Voronoi
diagram on the surface.

The authors describe their algorithm for computing this representation
as a ``continuous Dijkstra'' algorithm because it proceeds in a
fashion analogous to Dijkstra's algorithm for shortest paths in a
graph.  The MMP algorithm proceeds by maintaining in a priority queue
the shortest yet-known paths to a number of intervals.  At each step
it removes the nearest interval in the queue, identifies the shortest
yet-known path as an actual shortest path to at least the nearest
point in the interval, and propagates the paths through the interval
to one or more opposite edges of the next face.  The output of the
algorithm is correct by an induction, and the runtime of the algorithm
is governed by a bound on the number of intervals it must visit and
propagate.

Two modifications are required in order to extend the MMP algorithm
to suit our purpose.  First, the definition of a Voronoi diagram in
\cite{Mou85} must be slightly modified: it includes a point $x$ in the
Voronoi cell of source $s$ if $x$ is as close to $s$ as to any other
source.  A better definition includes $x$ only if it has a unique
shortest path to $s$, shorter than any path to another source.  It is
straightforward that the Voronoi cells under the latter definition are
simply connected.
Because the algorithm concerns shortest paths to edge points only, and
the Voronoi diagram is constructed by standard planar techniques after
the MMP algorithm proper is complete, this modification requires no
change to the MMP algorithm itself.

Second, and more complex, both \cite{Mou85} and \cite{MMP} make an
assumption that the polyhedral surface is embedded in $\R^3$, or
at a minimum that a lesser property holds which is not satisfied by a general
polyhedral metric represented by a general triangulation.  The earlier
paper \cite{Mou85} explicitly disclaims such an assumption, but at the
outset of Section~3 it asserts that the restriction of a shortest path
to a face is a single line segment, which is indeed easy to see if the
triangulation derives from a Euclidean embedding or otherwise if each
edge is a unique shortest path, but is not true in general.  The use of this
assertion is in the complexity analysis, so we repeat the complexity
analysis after proving a relaxed version of the assertion, consisting
of a bound on the number of line segments that may make up the
restriction of a shortest path to a face.  This re-analysis makes up the
remainder of this section.

Our argument requires an adjustment to one concept used throughout the analysis
in~\cite{MMP}.  For an edge $e$ and a face $f$ bordered by $e$, the
original analysis describes the paths associated with the edge-face
pair $(e, f)$ as ``$f$-free'', leading from the source to $e$ while
never passing through $f$.  In our case where faces need not consist
of shortest paths, the same algorithm will consider paths which are
not always $f$-free, but they will be \term{$f$-facing} in the sense
that they lead into $f$ when extended through their endpoint at $e$.
The analysis in~\cite{MMP} goes through unchanged with this broader
definition, with the exception of the complexity analysis.  We
rehearse the latter with our modifications after supplying ourselves
with a series of geometric lemmas.

The main object of our geometric study in this \biglc\ will be the
\term{piecewise-geodesic loop}, or simply \term{loop}, a non-self-intersecting
closed path on $M$ consisting of finitely many geodesic segments.
By the Jordan curve theorem, each loop partitions the rest of $M$ into two
\term{sides}, and we will sometimes identify one side as the
``interior'' and the other as the ``exterior''.  By a \term{closed
  side} of a loop we mean one side together with the loop itself.

\begin{lemma}\label{lem:19}
If a piecewise-geodesic loop of length $L$ has at least
two vertices on each closed side, then $L = \Omega(\ell \eps_8)$.
\end{lemma}

\begin{proof}
We shall repeatedly ``cut'' an angle of the loop, as follows.
Draw a chord of the loop very close to the angle, so that it encloses
a triangle with no vertices inside, and scale up by a homothety until
the chord meets either a vertex or an angle; if it meets an angle,
continue sliding the other endpoint of the chord until it meets
another angle or the chord meets a vertex.  The essential feature of
the cutting process is that it always decreases the total length of
the loop (by the triangle inequality),
so that every loop we work with is of length at most $L$.

Now, either the loop passes through two vertices, or it passes through
one and has at least one vertex on each side, or it has two on each
side.  Clearly if it passes through two vertices it has length at
least $2\ell$ and we are done.  We shall first reduce the case of at
least two vertices on each side to the case of one vertex on the loop
and at least one on each side.

At least one side of the original loop has total defect at most
$2\pi$, so identify one such side as the ``inside''.
We shall move the loop consistently
toward the inside.  At least some angles of the loop open toward the
inside, unless the loop has no angles at all, in which case it is a
closed geodesic, the
surface is locally a cylinder, and we may slide the loop
perpendicularly until we hit a vertex and thereby create an angle.
Now we cut an angle that opens
toward the inside, which either eliminates a vertex from the interior,
making it an angle of the loop, or reduces the number of angles in the
loop.  Therefore repeating this finitely many times brings us to a
loop which contains just one vertex $v$ in its interior, passes
through at least one other vertex $u$, and still has length at most $L$.

Once we reach this stage, we proceed by cutting any angle which is not
the vertex $u$.  If any such cut hits a vertex $w$, then $u$ and $w$
are within distance $L/2$ and $L \geq 2\ell$.  Otherwise, we cut until
$u$ is the only angle remaining in the loop, and because we
encountered no vertices there is still only the one vertex $v$ in the
interior.  Therefore the interior is metrically a cone, and by
elementary geometry $L \geq 2 \mathop{\rm dist}(u,v) \sin(\pi-\delta_v/2) \geq
2\ell \sin(\eps_8/2) = \Omega(\ell \eps_8)$.
\end{proof}

Call a piecewise-geodesic loop \term{live} if it meets the condition
of Lemma~\ref{lem:19}, having at least two vertices on each side.

Consider a digon $xGyHx$, where $x$ and $y$ are points and $G$ and $H$
two geodesics.  By local geometry, the extensions of $G$ and $H$ through $x$ are
on the same side of the digon, as are the two extensions through $y$.
If the extensions through $x$ and through $y$ are on opposite sides,
we call $xGyHx$ an \term{inside-out} digon, and otherwise all four
extensions are on the same side and we call it a \term{normal} digon.

We will repeatedly use the following basic fact: on a polyhedral
metric a digon always encloses at least one vertex on each side.
Further, because by Gauss-Bonnet a normal digon encloses a total
defect less than $2\pi$, its closed exterior contains defect greater than
$2\pi$ and therefore at least two vertices.

\begin{lemma}\label{lem:20}
  If non-self-crossing geodesics $G$ and $H$ in $M$ form an inside-out
  digon $xGyHx$ with only one vertex on one side, then $G$ and $H$
  end on that side after crossing at most $O(1/\eps_8)$ times.
\end{lemma}
\begin{proof}
  Call the side with one vertex the interior, arbitrarily, and by
  renaming let $y$ be the vertex through which $G$ and $H$ enter the
  interior.  Extend $G$ into the interior until its next crossing $z$
  with $H$.

  If $z$ lies on $xHy$, then $xGzHx$ and $zGyHz$ are normal digons
  whose interiors partition the interior of $xGyHx$.  But each of
  these interiors must contain a vertex, making at least two vertices
  inside $xGyHx$, a contradiction.

  If $yGzHy$ is a normal digon, then it must enclose a vertex in its
  interior, and the region interior to $xGyHx$ and exterior to $yGzHy$
  has no vertices.  Now $G$ and $H$ continue into this region;
  extending $G$, it cannot cross itself, but if it crosses any segment
  of $H$ it divides this region into two pieces, of which one is a
  digon, which with no vertices is impossible.  So $G$ must end in
  this region, and similarly $H$ must end, and there is only one
  crossing $z$ in the interior of $xGyHx$.

  Otherwise $yGzHy$ is an inside-out digon, dividing the interior of
  $xGyHx$ into a quadrilateral region $xGyHzGyHx$ and a region bounded
  only by $yGzHy$, which we call the interior.  This interior is an
  entire side of $yGzHy$, so it must contain the vertex.

  We have one crossing for each successive inside-out digon that $G$
  and $H$ form, and at most one for a normal digon at the end of the
  geodesics.  Now consider the angles at which $G$ and $H$ cross at
  each successive crossing $x, y, z, \dotsc$.  The difference between
  successive angles is the total external angle of the interior of the
  digon, which equals $2\pi$ minus the total defect enclosed, which is
  at most $2\pi - \eps_8$.  Therefore successive crossing angles
  increase by at least $\eps_8$, so because each angle is less than
  $\pi$ we have at most $\pi/\eps_8$ inside-out digons before the
  paths end.
\end{proof}

The next lemma is the one employed in our extended complexity
analysis.

\begin{lemma}\label{lem:21}
  On the polyhedral metric $M$, a non-self-crossing geodesic $G$ of
  length $L \geq \ell$ and a shortest path $H$ can cross at most
  $O(L/\ell \eps_8)$ times.
\end{lemma}

\begin{proof}
  Consider the sequence along $G$ of its crossings with $H$.  We show
  that for each crossing $y$, either its two neighbor crossings $x$
  and $z$ are separated by a length at least $\Omega(\ell \eps_8)$ of
  $G$, or one side of $y$ along $G$ has at most $O(1/\eps_8)$
  crossings in total.  The desired bound follows.

  Let three consecutive crossings with $H$ along $G$ be $x, y,$ and
  $z$.  Then the two digons $xGyHx$ and $yGzHy$ partition $M$ into
  three regions.

  If either digon is inside-out, say $xGyHx$, then either it is live
  or it has only one vertex on one side.  In the latter case, by
  Lemma~\ref{lem:20} there are $O(1/\eps_8)$ crossings toward that
  side before $G$ and $H$ end, and we are done.

  Otherwise both digons are normal.  Therefore they each have at least
  two vertices in their exterior, so if either one has at least two
  vertices in its interior then it is live.  Otherwise they each have
  exactly one interior vertex, so because $M$ has at least four
  vertices there are two vertices in the remaining region and its
  boundary, either the quadrangle $xGyHzGyHx$ or the digon $xGzHx$ if
  $xyz$ are out of order on $H$, is live.

  Now we have a live digon, or a live quadrangle whose edge set is a
  union of digons.  By Lemma~\ref{lem:19} the live digon or quadrangle
  has length $\Omega(\ell \eps_8)$.  Each digon consists of one
  segment from $G$ and one from $H$, so since $H$ is a shortest path
  at least half the total length of the digon must be in $G$.
  Therefore we have at least a length $\Omega(\ell \eps_8)$ in either
  $xGy$, $yGz$, or their union, so that the length of the segment
  $xGz$ is at least $\Omega(\ell \eps_8)$ as claimed.
\end{proof}

\comment{
\begin{lemma}
  Let $G$ be a non-self-crossing geodesic on $M$ with endpoints $x$
  and $y$, and let $s$ be an arbitrary point on $M$.  Let $P_x$ and
  $P_y$ be the shortest paths from $s$ to $x$ and $y$ respectively.
  Then the cycle $sP_xxGyP_ys$ winds around any point of $M$ at most
  $O(L/\ell \eps_8)$ times.
\end{lemma}
\begin{proof}

\end{proof}
}

Now we proceed to the argument of~\cite{MMP}, making the necessary
extension to the complexity analysis of the continuous Dijkstra
algorithm.

\begin{lemma}\label{lem:22}
  At the conclusion of the continuous Dijkstra algorithm,
  each edge-face pair $(e, f)$ has at most $O(n^2 S / \eps_8)$
  intervals in its interval list.
\end{lemma}
\begin{proof}
  We follow the proof of Lemma 7.1 in \cite{MMP}, extending it where
  necessary.  First, consider a single source $v$, and let $I_1, I_2,
  \dotsc, I_K$ be the interval list for the edge-face pair $(e, f)$
  that would be produced from the single source $s$.
  Let $x_i$ be a point interior to interval
  $I_i$, and let $x_0$ and $x_{K+1}$ be the endpoints of $e$.  As in
  \cite{MMP}, the intervals are ordered so that $f$ is on the left as
  we pass from $I_j$ to $I_{j+1}$, and we should name $x_0$ and
  $x_{K+1}$ so that $f$ is on the left as we walk from $x_0$ toward
  $x_{K+1}$.  Draw the shortest $f$-facing paths from $s$ to each
  $x_i$, calling them $P_i$.  Now between each $P_i$ and $P_{i+1}$ we
  have a region $sP_ix_iex_{i+1}P_{i+1}s$, the region on the right as
  we traverse that cycle.  Clearly there is a vertex in the interior
  of that region, or else $P_i$ and $P_{i+1}$ would intercept the same
  sequence of edges so that $x_i$ and $x_{i+1}$ would belong to the same
  interval.  But the whole cycle $sP_0x_0ex_{K+1}P_{K+1}s$ can wind
  about any particular vertex $v$ at most $O(D / \ell \eps_8)
  = O(S / \eps_8)$ times,
  because a shortest path from $v$ to a point on the outside will
  cross $e$ only that many times by Lemma~\ref{lem:21}.  Therefore at
  most $O(S / \eps_8)$ of the $K-1$ such regions may enclose any
  particular vertex, and so $K = O(n S / \eps_8)$.

  Now to complete the proof, we study the actual interval list
  produced by the algorithm when it runs with all our sources.  Each
  interval $I$ is a contiguous region on which the shortest paths lead
  from a particular source $s$ through a particular sequence of edges.
  Suppose that two intervals $I, I'$ derived from the same source and
  sequence of edges, so that some other interval $J$ intervened which
  derived from another source $t$, or from $t = s$ but with a different
  sequence of edges.  Then we may place $s$ and $t$ into the plane of
  $f$ by edge-unfolding along the respective sequences of edges, so
  that the points of $I$ and $I'$ are closer to $s$ than to $t$ but
  the points of $J$ are closer to $t$ than to $s$.  But this is
  impossible because our distances are decisive.  Therefore at most
  one interval in the actual interval list derives from a given source
  and sequence of edges, so the interval list contains at most one
  interval for each interval in the per-source interval lists, for a
  total of $O(n^2 S / \eps_8)$ intervals.
\end{proof}

\begin{lemma}\label{lem:23}
  The continuous Dijkstra algorithm runs in time $\tw O(n^3 S /
  \eps_8)$.
\end{lemma}
\begin{proof}
  By Lemma~\ref{lem:22} there are at most $O(n^3 S / \eps_8)$
  intervals in the algorithm's entire final list.  Each call to
  \textsc{Propagate} causes one final interval, and creates at most
  two new intervals.  Therefore there are $O(n^3 S / \eps_8)$
  intervals ever created.  The ordered-set operations required on
  interval lists take time $O(\log (n S / \eps_8))$, there are
  $O(1)$ of these and $O(1)$ other work in each round of the
  algorithm, one round per final interval, and so the total time is
  $O(n^3 (S / \eps_8) \log(n S / \eps_8)) = \tw O(n^3 S /
  \eps_8)$ as claimed.
\end{proof}

\smallsec{Unweighted Delaunay triangulation on a polyhedral surface}
\label{subsec:unweighted-Delaunay}

Next we give an efficient algorithm for computing an unweighted Delaunay
triangulation on a polyhedral surface.  An algorithm based on
successively flipping edges which fail the local convexity condition
was previously known to terminate, but is not believed to finish in
polynomial time \cite{Indermitte+,Rivin}.
Instead, here we use the unweighted Voronoi diagram computed from the previous
section.  Note that the continuous Dijkstra algorithm can be modified to
compute weighted Voronoi (power) diagrams, but it seems difficult to
transform such a diagram into a corresponding weighted Delaunay triangulation;
thus we focus on the unweighted case for now.

\begin{algorithm}{Polyhedral-Delaunay}
Begin by computing the unweighted Voronoi diagram using the generalized MMP
algorithm described in \smallref\ \ref{subsec:shortest-paths}.  For each two
Voronoi cells $x$ and $y$ that touch, and each segment $e$ on their mutual
boundary, take a point $p$ on that segment.  Because $p$ is on the two
Voronoi cells, the disk $D_p$ of radius $px = py$ about $p$ has no vertices
in its interior, so it is isometric to a planar disk.  Let $e'$ be
the segment $xy$ through $D_p$.  It is straightforward to see that
$e'$ does not depend on the choice of $p$.

{From} the data provided by the MMP algorithm at $p$,
we may compute in constant time the length of the
radii $px$ and $py$ of $D_p$ and the angles they make at $p$, $x$, and
$y$, so by trigonometry we may also compute in constant time the
length of $e'$ and the angles it makes at $x$ and $y$.  Let $T$
consist of the segments $e'$ for each Voronoi boundary segment $e$,
described by their lengths and their ordering about each vertex.

If the Voronoi diagram has any points $p$ at which more than three
cells meet, then the disk $D_p$ about $p$ has the source of each cell
on its boundary and no vertices in the interior.  The chords between
adjacent vertices on the boundary of $D_p$ are already present in $T$
as the segments derived from the Voronoi edges meeting at $p$, and we
add to $T$ further chords chosen to triangulate arbitrarily the
polygon they form.
\end{algorithm}

\begin{lemma}
  Algorithm \algo{Polyhedral-Delaunay} computes an unweighted
  Delaunay triangulation in $O(n)$ time plus the
  $\tw O(n^3 S / \eps_8)$ time to compute the unweighted Voronoi diagram.
\end{lemma}
\begin{proof}
  In the conversion from Voronoi to Delaunay, we spend $O(1)$ for each
  Voronoi edge, and $O(d)$ work for each Voronoi vertex of degree~$d$,
  for a total of $O(n)$ work.

  For correctness, we need to show that the edges of $T$ form a
  triangulation, and that the triangulation is locally convex at each
  edge.

  First, we show that the edges of $T$ do not intersect.  Suppose that
  edges $e'$ and $f'$, which were respectively drawn through disk
  $D_p$ from $A$ to $C$ and through $D_q$ from $B$ to $D$, intersect
  at $X$.  Now $D_p$ contains no vertices in its interior, so $f'$
  must extend from $X$ at least to the boundary of $D_p$ in each
  direction, and by the classical theorem on the power of a point, $XB
  \cdot XD \geq XA \cdot XC$.  Similarly $XA \cdot XC \geq XB \cdot
  XD$, so the two powers are equal and $D_p = D_q$ is a disk with $A$,
  $B$, $C$, and $D$ all on its boundary.  Consequently $p = q$ is a
  point at which at least the four cells for $A$, $B$, $C$, and $D$
  meet, and $e'$ and $f'$ are diagonal chords chosen in the final step
  of the algorithm.  But these chords are chosen as a triangulation,
  so they do not intersect.

  Now, we count the edges of $T$ and apply Euler's formula to deduce
  that they form a triangulation.  The Voronoi diagram has $n$ faces;
  let it have $m_1$ edges and $m_0$ vertices, so that $m_1 - m_0 = n -
  2$ by Euler's formula.  Each Voronoi vertex of degree $d$
  contributes $d - 3$ edges in the last step of the algorithm, so a
  total of $2m_1 - 3m_0$ edges come from this step, for a total of
  $3m_1 - 3m_0 = 3n - 6$ edges in $T$.  Since $T$ has $n$ vertices, by
  Euler's formula it must have $2 + (3n-6) - n = 2n - 4$ faces.
  Therefore the average degree of a face is $2\cdot(3n-6) / (2n-4) =
  3$.  Since $M$ is connected, every set of Voronoi cells must share
  edges with its complement, so $T$ is connected, and therefore by
  geometry it can have no digons or empty loops.  Therefore every face
  of $T$ is a triangle.

  Finally, consider an edge $e'$ of $T$ separating triangles $ACD$ and
  $CAB$.  In the unweighted case, the local convexity condition
  reduces to a requirement that when $ACD$ and $CAB$ are developed
  together into the plane, $D$ is not in the interior of the
  circumcircle of $CAB$.  But in this development, $B$ and $D$ lie on
  opposite sides of line $AC$, and the disk $D_p$ through which $e'$
  was drawn touches $A$ and $C$ and contains neither $B$ nor $D$ in
  its interior.  Therefore the circumcircle of $CAB$ has its center at
  least as far to the $B$ side of $AC$ as $D_p$ does, and contains a
  subset of those points to the $D$ side of $AC$ that $D_p$ does, so
  that it does not contain $D$ in its interior and $e'$ is locally convex.
\end{proof}

\smallsec{Weighted Delaunay triangulation on a polyhedral surface}
\label{subsec:weighted-Delaunay-algo}

Algorithm \algo{Polyhedral-Weighted-Delaunay} finds a weighted
Delaunay triangulation for a polyhedral surface $M$ with vertex
weights $w$ satisfying certain conditions.  The weights $w$ must obey
a system of linear inequalities that guarantee that any edge $AC$
which is not strictly locally convex lies in a convex quadrilateral
consisting of two neighboring triangles $ABC$ and $CDA$, and therefore
can be ``flipped'' to substitute the contrary diagonal $BD$.
Proposition 4 of \cite{BI} guarantees that every $w$ arising from a
generalized convex polyhedron, which includes every $w$ we consider,
satisfies this condition.

The algorithm makes use of the
following geometric observation.  In
polyhedral surface $M$ with vertex weights $w$ and triangulation
$T$, let the \term{height} function on $M$ be the unique function $h :
M \to \R$ that within each triangle of $T$ is quadratic with unit
quadratic part along every line segment, and such that at each vertex
$v$ the height
agrees with the weight, $h(v) = w(v)$.  We call this function the
``height'' because when the vertex weights arise from a radius
assignment $r$ , it coincides with the squared distance from the
apex in the tetrahedra built from $T$ and $r$.  (In this setting, $h$ is the
function $q_{T,r}$ of \cite{BI}.)
Then we have the following lemma,
which is mentioned in \cite{BI} without proof.

\begin{lemma}\label{lem:25}
  If $T$ is not locally convex at edge $xz$ with neighboring triangles
  $xyz$ and $wxz$, then replacing $xyz$ and $wxz$ with $xyw$ and $wyz$
  strictly increases the height of the points in the interior of $wxyz$.
\end{lemma}

\begin{proof}
  Consider the points in the intersection
  $U$ of triangles $xyz$ and $xyw$.  This is without loss of
  generality, because every point inside $wxyz$ lies in at least one
  equivalent such intersection.
  Develop both triangles into the
  plane, and for clarity of discussion, orient the diagram so that
  edge $xy$ is horizontal, and the alternate vertices $w, z$ are above
  $xy$.  (Because $xz$ is not locally convex, $wxyz$ is a convex
  quadrilateral so that $w$ and $z$ are necessarily on the same side
  of $xy$.)

  The height function on segment $xy$ is determined in both
  triangulations as the unique quadratic function with unit quadratic
  part along $xy$ and agreeing with the vertex weights at $x$ and
  $y$.  The height function on $U$ is easily seen to be determined by
  its gradient at $xy$, and the gradient of $h$ at any point in a
  triangle $v_iv_jv_k$ is twice the displacement vector from the
  center $C(v_iv_jv_k)$.  Both centers $C(xyz)$ and $C(xyw)$ lie
  on the radical axis $\axis(x, y)$ of $x$ and $y$, the line on which
  $\pi_x(p) = \pi_y(p)$, which is perpendicular to $xy$.  Therefore,
  the height function at any point in $U \setminus xy$ is greater in
  $xyz$, greater in $xyw$, or equal in both triangles just if
  $C(xyz)$ is lower than $C(xyw)$ along $\axis(x, y)$, is higher than
  $C(xyw)$, or coincides with $C(xyw)$ respectively.

  But if $T$ with triangles $xyz$ and $wxz$ fails to be locally convex
  at $xz$, then $\pi_w(C(xyz)) < \pi_x(C(xyz))$, so that $C(xyz)$
  falls on the $w$ side of the radical axis $\axis(x, w)$.  Because
  $w$ is above $x$ and $\axis(x, w) \perp xw$, the $w$ side is the
  upper side, so that $C(xyz)$ falls above the intersection of
  $\axis(x, w)$ with $\axis(x, y)$, which is $C(xyw)$.  Therefore the
  height function is greater in triangle $xyw$ than in triangle $xyz$,
  as required.
\end{proof}

Let $XY$ be an edge of the triangulation, and let $P$ be on $XY$.
Then
$$ h(P) = \frac{PY \cdot w(X) + PX \cdot w(Y)}{XY} - PX \cdot PY . $$
Therefore in a pair of triangles $WXY$, $WYZ$ forming a quadrilateral
$WXYZ$ with $P = \overline{WY} \cap \overline{XZ}$, if $WY$ is in the
triangulation we have
$$ h(P) = \frac{PY \cdot w(W) + PW \cdot w(Y)}{WY} - PW \cdot PY $$
and if $XZ$ is in the triangulation we have
$$ h(P) = \frac{PZ \cdot w(X) + PX \cdot w(Z)}{XZ} - PX \cdot PZ $$
so that for any assignment of weights to $W$, $X$, $Y$, and $Z$ we may
determine by a comparison of these quantities whether triangles $WXY$
and $WYZ$ would flip to $XYZ$ and $XZW$ or vice versa.  In particular,
consider $w(X)$, $w(Y)$, and $w(Z)$ as fixed.  Then $WXY$ and $WYZ$
are preferred over $XYZ$ and $XZW$ or vice versa just if $w(W)$ is
greater than or less than
\begin{equation}\label{eq:flip}
t_{W,XYZ} \defeq \frac{WY}{PY} \left(
 \frac{PZ \cdot w(X) + PX \cdot w(Z)}{XZ} - \frac{PW \cdot w(Y)}{WY}
  - PX \cdot PZ + PW \cdot PY
\right).
\end{equation}

Now, suppose we have a weighted Delaunay triangulation $T$ for some
vertex weights $w$, and consider a new vertex weighting $w'$ which
differs from $w$ at only one vertex $v$, with $w'(v) < w(v)$.  The
local convexity criterion is identical in $w'$ and $w$ for all pairs
of adjacent triangles neither of which is incident to $w$.  Further,
by our discussion of equation~\ref{eq:flip}, any edge in $T$ which
lies across a triangle from $v$ remains locally convex under $w'$
because the weight of $v$ has only declined.

The following algorithm \algo{Reweight} makes use of $T$ to compute
a Delaunay triangulation $T'$ for $w'$.

\begin{algorithm}{Reweight}
Let $v$ have degree $d$.  For each three consecutive neighbors $x, y,
z$ of $v$ in $T$, define $t_y \defeq t_{v,xyz}$.  Compute the $d$
values $t_y$ using equation~\ref{eq:flip} and store them in a
max-priority queue $Q$.  Now repeat the following steps until done:

Remove the maximum element $t_y$ from $Q$.  If $t_y \leq w'(v)$, we
are done.  Otherwise, replace edge $vy$ with $xz$.  Recompute $t_x$
and $t_z$ to reflect the modified triangulation, and update the
priority queue.

When the iteration is done, the resulting triangulation is $T'$.
\end{algorithm}

\begin{lemma}\label{lem:27}
  Algorithm \algo{Reweight} converts a Delaunay triangulation for $w$
  into a Delaunay triangulation for $w'$ in time $O(n \log n)$.
\end{lemma}
\begin{proof}
  There are $d < n$ elements in the priority queue, each of which is
  inserted once and removed at most once, and there are two updates
  and one operation on $T$ for each removal.  This totals $O(d)$
  operations which may each be done in time $O(\log d)$, for a total
  time $O(d \log d) = O(n \log n)$. 

  We prove correctness by induction.  Let \algo{Reweight} change $m$
  edges, let $w(v) = t_0 \geq t_1 \geq \dotsb \geq t_{m+1}$ with $t_i$
  the value removed from $Q$ in step $i$, and let $T = T_0, T_1,
  \dotsc, T_m = T'$ be the successive triangulations considered.
  Write $w_t$ for the vertex weighting that differs from $w$ only in
  setting the weight of $v$ to a value~$t$.  We claim that for $i = 0,
  \dotsc, m$, triangulation $T_i$ is Delaunay for weightings $w_t$
  with $t_i \geq t \geq t_{i+1}$. 

  For $i = 1, \dotsc, m$, triangulation $T_{i-1}$ is Delaunay for
  $w_{t_i}$ by hypothesis.  Triangulation $T_i$ differs only in one
  pair of edges, at which by construction it remains locally convex
  for $w_{t_i}$.  Now by the discussion above, the only edges in $T_i$
  that may fail local convexity under any $w_t$ with $t \leq t_i$ are
  those incident to $v$.  But $t_{i+1}$ was constructed as the maximum
  weight for $v$ at which the local convexity condition would reach
  equality for any of these edges.  Therefore $T_i$ remains Delaunay
  for weightings $w_t$ with $t_i \geq t \geq t_{i+1}$ as claimed.

  In the base case of $i = 0$, triangulation $T_0 = T$ is Delaunay for
  $w$ by precondition.  Then $T_0$ remains Delaunay for weightings
  $w_t$ with $t_0 \geq t \geq t_1$ by the same argument as in the
  inductive case, and the induction is complete.

  Now setting $i = m$, we find that $T' = T_m$ is Delaunay for
  weightings $w_t$ with $t_m \geq t \geq t_{m+1}$.  But by the
  termination condition, $t = w'(v)$ lies in this range.  Consequently
  $T'$ is Delaunay for $w_{w'(v)} = w'$ as required.
\end{proof}

With \algo{Polyhedral-Delaunay} and \algo{Reweight} as subroutines, it
is now straightforward to
compute a weighted Delaunay triangulation for any vertex weighting
$w$ on a polyhedral surface $M$.

\begin{algorithm}{Polyhedral-Weighted-Delaunay}
First, we compute an unweighted Delaunay triangulation $T_0$ on $M$ by
Algorithm \algo{Polyhedral-Delaunay}.
Then we adjust the weights.  The local convexity condition is
unchanged by an additive constant on all the vertex weights, so $T_0$
is Delaunay for a vertex weighting $w_0$ with $w_0(v) = \max_u w(u)$
for all $v$.  Number the vertices $v_1, \dotsc, v_n$ in arbitrary
order, let vertex weighting $w_i$ coincide with $w$ on vertices $v_j$
with $j \leq i$ and with $w_0$ elsewhere, and apply
algorithm~\algo{Reweight} to compute triangulations $T_1, T_2,
\dotsc, T_n$ in turn that are Delaunay for weightings $w_1, \dotsc,
w_n$.  Then $T = T_n$ is our result.
\end{algorithm}

\begin{lemma}
  Algorithm \algo{Polyhedral-Weighted-Delaunay} computes a Delaunay triangulation for $w$ in
  $O(n^2 \log n)$ time, plus the $\tw O(n^3 S / \eps_8)$ time to compute
  an unweighted Voronoi diagram.
\end{lemma}
\begin{proof}
  Algorithm \algo{Polyhedral-Delaunay} costs $\tw O(n^3 S / \eps_8)$ time to
  compute an unweighted Voronoi diagram via MMP, plus $O(n)$ time to
  convert into an unweighted Delaunay triangulation $T_0$.
  We then apply \algo{Reweight} $n$ times, in time $O(n \log n)$ each,
  for $O(n^2 \log n)$ total additional time.

  The output $T = T_n$ is correct by a simple induction using
  Lemma~\ref{lem:27}.
\end{proof}

\label{lem:4}

\subsubsection*{Acknowledgments.}
We thank Jeff Erickson and Joseph Mitchell for helpful discussions about
shortest paths on non-shortest-path triangulations, and the anonymous
referees for helpful comments.

\bibliography{references} \bibliographystyle{halpha}